\documentclass[conference]{IEEEtran}
\IEEEoverridecommandlockouts

\usepackage{makecell}
\usepackage{tcolorbox} 
\usepackage{array}
\usepackage{longtable}
\usepackage{comment}
\usepackage{enumitem}
\usepackage{booktabs}
\usepackage{hyperref}
\usepackage{float}
\usepackage{amsmath}
\usepackage{mathpazo}
\usepackage{datetime}
\usepackage{color, colortbl}
\usepackage{adjustbox}

\usepackage{cite}
\usepackage{amsmath,amssymb,amsfonts}
\usepackage{algorithmic}
\usepackage{graphicx}
\usepackage{textcomp}
\usepackage{xcolor}
\def\BibTeX{{\rm B\kern-.05em{\sc i\kern-.025em b}\kern-.08em
    T\kern-.1667em\lower.7ex\hbox{E}\kern-.125emX}}

\definecolor{VeryHigh}{rgb}{0.90,0.44,0.43}
\definecolor{High}{rgb}{0.92,0.59,0.47}
\definecolor{Medium}{rgb}{0.94,0.72,0.51}
\definecolor{Low}{rgb}{0.68,0.81,0.52}
\definecolor{VeryLow}{rgb}{0.47,0.73,0.50}  

\definecolor{Blue}{rgb}{0.4,0.7,0.8}

\definecolor{Grey}{rgb}{0.85,0.85,0.85}

\usepackage{bookmark}
\usepackage{multirow}
\usepackage{xspace}
\usepackage{tabularx}

\usepackage{url}
\usepackage{todonotes}
\usepackage{listings}

\newcommand{\thickhline}{%
    \noalign {\ifnum 0=`}\fi \hrule height 1pt
    \futurelet \reserved@a \@xhline}
\newcolumntype{"}{@{\hskip\tabcolsep\vrule width 1pt\hskip\tabcolsep}}

\definecolor{bottleGreen}{RGB}{13, 172, 91}
\def\BibTeX{{\rm B\kern-.05em{\sc i\kern-.025em b}\kern-.08em
    T\kern-.1667em\lower.7ex\hbox{E}\kern-.125emX}}

\begin{document}

\title{Evaluating the Security of Open Radio Access Networks}

\author{

    \IEEEauthorblockN{
        Dudu Mimran\textsuperscript{1},
        Ron Bitton\textsuperscript{1}, 
        Yehonatan Kfir\textsuperscript{1}, 
        Eitan Klevansky\textsuperscript{1}, 
        Oleg Brodt\textsuperscript{1}, \\ 
        Heiko Lehmann\textsuperscript{2}, 
        Yuval Elovici\textsuperscript{1}, and 
        Asaf Shabtai\textsuperscript{1}
    }
    
    \IEEEauthorblockA{
        Ben-Gurion University of the Negev\textsuperscript{1} \\
        Deutsche Telekom AG, T-Labs\textsuperscript{2} 
    }
}

\maketitle

\begin{abstract}
The Open Radio Access Network (O-RAN) is a promising RAN architecture, aimed at reshaping the RAN industry toward an open, adaptive, and intelligent RAN.
In this paper, we conducted a comprehensive security analysis of Open Radio Access Networks (O-RAN).
Specifically, we review the architectural blueprint designed by the O-RAN alliance -- A leading force in the cellular ecosystem.
Within the security analysis, we provide a detailed overview of the O-RAN architecture; present an ontology for evaluating the security of a system, which is currently at an early development stage; detect the primary risk areas to O-RAN; enumerate the various threat actors to O-RAN; and model potential threats to O-RAN.
The significance of this work is providing an updated attack surface to cellular network operators.
Based on the attack surface, cellular network operators can carefully deploy the appropriate countermeasure for increasing the security of O-RAN.
\end{abstract}


\section{\label{sec:into}Introduction}
In recent years, the number of cellular networks users has increased dramatically. At the end of 2017, 4.8 billion unique mobile subscribers were representing 65\% of the world’s population \cite{global2017state}. By the end of 2021, the volume of cellular data traffic produced by mobile devices it is expected to grow by 47\% (annual growth rate) \cite{ciscowebsite}. In addition to providing services to mobile devices, cellular networks support other diverse applications \cite{parkvall2017nr}, such as real-time video, connected autonomous cars, distributed sensors, and smart manufacturing. The new use cases and applications pose new requirements for cellular networks; where some require high bandwidth with high latency (e.g., video streaming), and others require low bandwidth and low latency (e.g., connected cars)—requirements which push the cellular networks to become more adaptive and capable. To meet the new requirements, new proposed cellular architectures were introduced aiming at giving the radio access network (RAN) \cite{bonati2020open} a more prominent role, which was traditionally responsible mainly for transmitting data from the user equipment (UE) to the core network for further processing. A promising RAN architecture that seems to get adopted worldwide is the one suggested by the O-RAN Alliance \cite{alliance2021ran}.
 
The O-RAN architecture is based on open specifications and disaggregation, breaking the RAN into multiple open and capable units, leveraging cloud technologies to achieve scalability and reliability concerning the number of connections to the cellular network. Furthermore, O-RAN introduces new paradigms such as an extensible RAN where third-party applications and services can be integrated into the platform to evolve the capabilities in an agile manner. Another key design consideration in O-RAN is the usage of machine learning for efficiently managing the network resources across the different applications and services - an example application for an ML capability in O-RAN is the ability to manage multiple services with different quality of service (QoS) on the same network where the decisions on how to allocate the resources at real-time across the services are decided autonomously. The openness of the architecture, the introduction of new IT technologies as well as machine learning into the RAN offer great promise in terms of meeting the requirements \cite{parvez2018survey}. However, those architectural changes introduce dramatic changes to the attack surface of the RAN.
 
This paper presents a systematic threat analysis of the O-RAN architecture that is unlike the common approach for conducting security evaluation for existing technologies. We have evaluated design blueprints and documents and not a mature technology/implementation stack and for that goal, we created a specialized methodology. We have decided not to analyze the reference code of O-RAN due to the fact it does not represent a complete implementation of O-RAN and is a basis for rapid changes as the O-RAN specifications evolve and its adoption grows.  Furthermore, we have not covered different relevant countermeasures to the identified threats as the analysis is done at the conceptual level and concrete countermeasure selection is not relevant at this stage. To the best of our knowledge, such threat analysis on the architecture level and transferable risks from other domains has not been performed in the past, and it aims to provide deep insight for the designers and implementers (operators) of the O-RAN concept, assisting in risk assessment and mitigation planning. The importance of such an early assessment of security is rooted in the concept of security by design where the effectiveness of iterating an architecture in its early stages towards a secure design is in magnitude more effective and results in more robust results vs. conducting the security analysis at later stages in a technology lifetime. Furthermore, the analysis in this paper can serve as a baseline for operators in building cyber defense strategies for protecting O-RAN networks, including risk assessments and a guideline for which countermeasures to deploy.

The contributions of this paper are as follows:
\begin{enumerate}[leftmargin=*]
    \item We developed a complete security evaluation process for evaluating the O-RAN architecture including the following elements: an ontology specifically designed for evaluating O-RAN’s architecture security risks. The proposed ontology was inspired by the general methodology presented by NIST for modeling enterprise security risks \cite{singhal2012security}. Based on the ontology we have devised a taxonomy based on O-RAN architecture and the analysis goals. We also explain our threat survey methodology. This methodology can be reused for later evaluations of the next iterations in O-RAN architecture.
    \item Using our evaluation process we conducted a comprehensive general threat analysis of O-RAN. We surveyed past attacks in different domains, evaluated their applicability to the different risk areas within O-RAN and mapped their operational and security impact.
    \item Based on the threat analysis, we identified several key directions for future research for improving the overall security of the O-RAN architecture. 
\end{enumerate}

\section{\label{sec:background}The O-RAN Architecture}
\subsection{RAN Architecture Evolution} 
\begin{figure}[t]
    \centering
    \includegraphics[width=0.5 \textwidth]{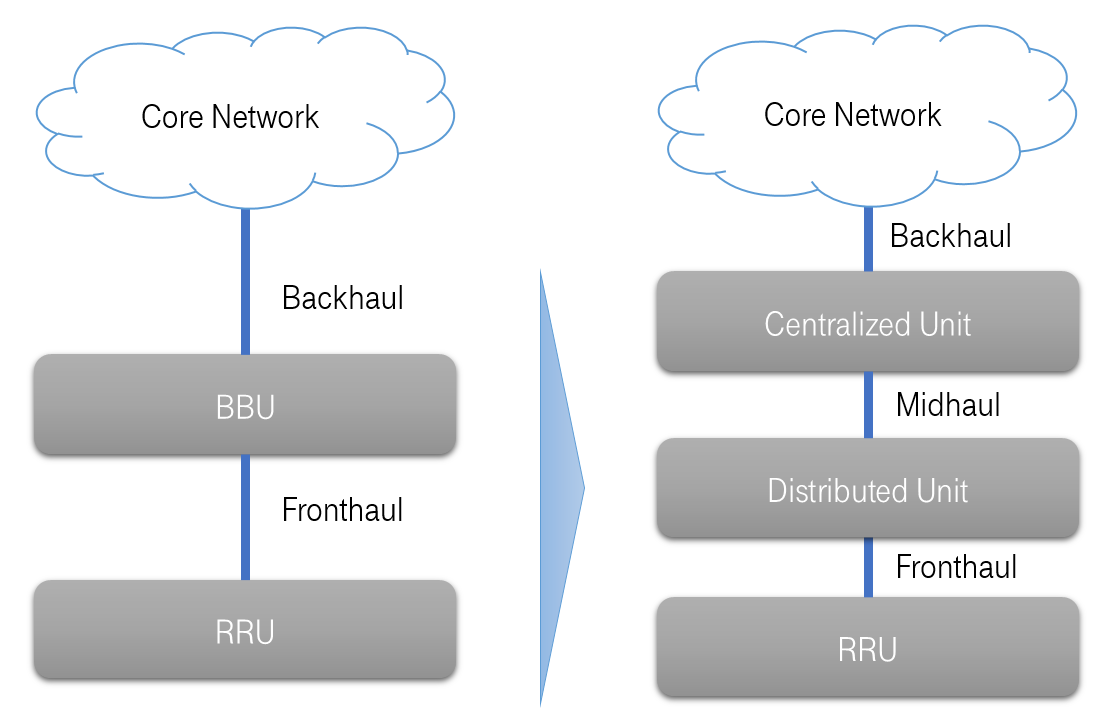}
    \caption{The transition to disaggregated RANs.}
    \label{fig:Disaggregated-RAN}
\end{figure}

Cellular infrastructure is made up of two types of networks (see Figure~\ref{fig:Disaggregated-RAN}): the radio access network (RAN) and the core network. 5G is the new radio access technology and a next-generation network architecture defined by the 3GPP (3rd generation partnership project), a coalition of telecommunications standard development organizations that create technical standards and specifications for cellular telecommunications technologies. The RAN provides wireless connectivity to mobile devices and acts as the final link between the cellular network and the user equipment (UE), such as a smartphone or connected car. The UE uses the new radio air interface, the radio-frequency (RF) portion in the circuit between the user equipment and a base station (usually frequency, channel bandwidth, and modulation scheme) for communication with RAN antennas, receiving the RF signals, and connecting the UE with the core network services via a transport network. The core network provides a wide range of services, such as call routing, user authentication, billing, etc. The 5G NR specification is subdivided into two frequency bands, FR1 (below 6 GHz, sharing the spectrum with 4G) and FR2 (mmWave, above 24 GHz ), which increases the amount of RF channels available to the network. 

A typical RAN is composed of a radio unit (RU) and a base station (containing baseband units, BBUs), whereas in the traditional architecture, the digital processing equipment was located near the radio unit antenna. In older mobile network generations (before 4G), the electronic equipment of the RU and BBU were coupled together at the bottom of the mobile antenna towers, and RF cables were used to connect the RU to the antennas at the top of the towers. However, this approach was inefficient in terms of signal performance and coping with changes in requirements, and eventually, the cellular industry transitioned the RU equipment to the top of the tower (connecting the base unit with fiber optics cables); hence the terms remote radio unit (RRU) and remote radio head (RRH). Traditional RAN (RRH hardware and BBU hardware and software) platforms are proprietary and commonly supplied by a single vendor. The interfaces between the RU and the BBU are defined by the vendor, and the applications running on them are tailored and optimized to the specific vendor’s equipment. Vendor-specific solutions allowed vendors to provide optimized, integrated solutions; however, it has been sub-optimal regarding the flexibility and scalability requirements of 5G networks. For example, introducing a new radio frequency band, adding faster interfaces, or supporting new emerging applications required upgrading the entire RAN. In addition, a proprietary RAN requires the vendor to develop all of the components, which increases both the cost of the RAN for operators as well operators’ dependence on specific vendors. 
Over 4G and 5G generations, the RAN architecture has evolved to become less centralized and more disaggregated (see Figure~\ref{fig:Disaggregated-RAN}) accompanied by a transition from the use of specialized equipment and software to the use of general-purpose hardware, virtualization, and adoption of cloud-native technologies. Disaggregation allowed deployment flexibility accommodating the complexity and diversity of 5G use cases.

Several RAN approaches were introduced over the years. One interesting approach for designing radio access networks is known as C-RAN, where the base station’s digital processing function is transferred to a regional cloud/edge data center. This approach has a very high communication overhead for maintaining low latency over the fronthaul link between the centralized site and the RUs. 
Another interesting approach was the vRAN (Virtualized RAN), which aimed to transition the base station to general-purpose hardware and run the RAN software in a virtual environment; the approach suffered in areas of low latency digital signal processing due to virtualization overhead. Both C-RAN and vRAN simplified maintenance and reduced costs dramatically, however, in both architectures, the dependency on a single vendor remained with all its disadvantageous. The major contributor to the effort of decoupling the implementation from the design which promotes disaggregation is the O-RAN Alliance \cite{alliance2021ran}. The O-RAN Alliance promotes a general-purpose and vendor-independent RAN solution by compartmenting the functional areas in the RAN and specifying open interfaces between the RAN components. For example, once the interface between the BBU and RU is defined in an open specification, it allows operators to use equipment from different vendors, thereby supporting various deployment scenarios based on various integration solutions. The combined capabilities of C-RAN / vRAN and open RAN specifications enable maximum flexibility in light of the demanding requirements. Such a combination is manifested in the O-RAN architecture, which is evolving to become the leading reference architecture for 5G RAN. Therefore, in this paper, we focus on the O-RAN architecture. 

\subsection{5G RAN Requirements}

The fifth-generation cellular network (5G) is expected to integrate and support modern environments seamlessly, yielding an interactive user-oriented information ecosystem \cite{jiang2017overview,liu20165g}. The main drivers for 5G are increased capacity and faster data rates and the need to service diverse ‘vertical’ industries with ultra-reliable and low-latency connectivity \cite{norp20185g}. The 5G performance requirements are as follows \cite{jiang2017overview,alnoman2017towards,mattisson2018overview}: 
\begin{itemize}
    \item \textbf{Low Latency:} less than 1 ms round-trip network delay latency.
    \item \textbf{Bandwidth:} data rates exceeding 10 Gb/s and capacity expansion (data traffic) by a factor of 1000.
    \item \textbf{Connection Density:} connectivity expansion by a factor of 100 (connected devices per square kilometer).
    \item \textbf{Energy Efficiency:} improve energy efficiency by a factor of 1000.
    \item \textbf{Mobility:} support mobility of over 500 km/hour.
    \item \textbf{Cost Reduction:} significant deployment costs reduction, by a factor of 10. 
\end{itemize}

The following three major 5G capabilities cover a wide range of use cases, applications, and scenarios \cite{ateya2018study}: 

\noindent \textbf{Enhanced mobile broadband (eMBB):} bandwidth-demanding use cases such as entertainment and media, live sport, online gaming, AR/VR, UltraHD, etc., the focus in these use cases is on higher data bandwidth capacities, latency improvements, and overall user experience enhancement. 

\noindent \textbf{Massive machine-type communications (mMTC):} the main driver for this capability is the connectivity and networking of a massive number (billions) of machines (IoT) to the cellular network. 

\noindent \textbf{Ultra-reliable and low-latency communications (uRLLC):} a capability which serves services and use cases that pose stringent latency and reliability requirements, enabling mission-critical applications such as industrial automation, smart grids, remote surgery, and vehicle-to-everything (V2X). 

\subsection{New Architectural Concepts in O-RAN}

\noindent\textbf{Slicing:} Network slicing is the ability of 5G networks to partition the physical network into several virtual networks. It allows provisioning an end-to-end connectivity and data processing tailored to serve a specific business use case. Using slicing, 5G networks can meet the scalability and flexibility requirements \cite{rost2017network}. Slicing offers better resource isolation, where key parameters such as bandwidth, transport network, security, or access point density can be configured and tuned per business quality of service (QoS) requirements. It allows operators to offer network slices in service-level agreements (SLAs) \cite{habibi2018structure}. Slicing is an end-to-end capability, done both in the core and the RAN \cite{elayoubi20195g}. Key technologies enabling slicing are NFV and SDN.

\noindent\textbf{Architectural openness:} open RAN’s core concept is the open interfaces; once the BBU and the RRU/RRH interfaces are open, operators can mix and match radio and digital processing units from different vendors. The openness vision promotes general purpose and vendor-neutral hardware and software with open interfaces between all decoupled RAN components. Openness means multi-vendor interoperability, which requires the enhancement of 3GPP interfaces \cite{alliance2018ran} as well as the addition of more additional interfaces. 

\noindent\textbf{Cloud and Virtualization:} The radio band spectrum is a scarce resource. Radio workload multiplexing improves the utilization of resources and supports the required economics of densification \cite{alliance2020ran}. An overall cost reduction theme involves replacing proprietary hardware and software with off-the-shelf general-purpose hardware and software. The RAN is no exception, and O-RAN promotes cloudification of the RAN technology and overall shift to cloud-native technologies where network functions are virtualized and are controlled by software-defined networking. Cloudification facilitates flexible resource provisioning and enables the centralization of the RAN infrastructure and the reduction of operational costs \cite{alliance2020ran}. Cloudification of Radio Access Networks is manifested by the introduction of a multitude of technologies mostly popular in the enterprise public and private cloud worlds such as machine virtualization (VM), containers, containers orchestration frameworks, software-defined networking, network function virtualization, and others. Furthermore, the development practices of cloud-driven applications and services are introduced as well to the world of O-RAN where functions such as DevOps, continuous integration, and continuous delivery, will become an inseparable part of the O-RAN application development lifecycle. A side effect of working on cloud-driven platforms is the explosive usage of open-source software packages, serving a big part in the evolution of cloud technologies in general. Open-source software is introduced both in the tools used by developers, integrators, and operators as well as in the concrete applications, services, and general runtime software stacks. 

\noindent \textbf{Machine Learning:} 5G architecture’s complexity, breadth of distributed and disaggregated components, and the required flexibility for supporting a plethora of innovative use cases and applications pose new cellular network management challenges. Network tasks such as optimization, deployment, orchestration, and operation are becoming increasingly complicated, to the point that human or rule-based management approaches are rendered ineffective. To cope with the growing complexity and required short time for decision making, machine learning-based capabilities are planned. O-RAN introduces system-level machine learning capabilities in the management aspect of the network as well as general-purpose support for third-party machine learning applications within the RAN architecture \cite{alliance2020ran}.

\subsection{O-RAN Architecture, Components, and Protocols}
Figure~\ref{fig:high_level_arch} provides a high-level view of the O-RAN architecture \cite{alliance2021ran}. The mobile network comprises multiple management domains such as core management, transport management, end-to-end slice management, etc. The Service Management and Orchestration (SMO) is responsible for managing the RAN domain, and it includes the following main parts: 

\begin{figure}
    \centering
    \includegraphics[width=0.5 \textwidth]{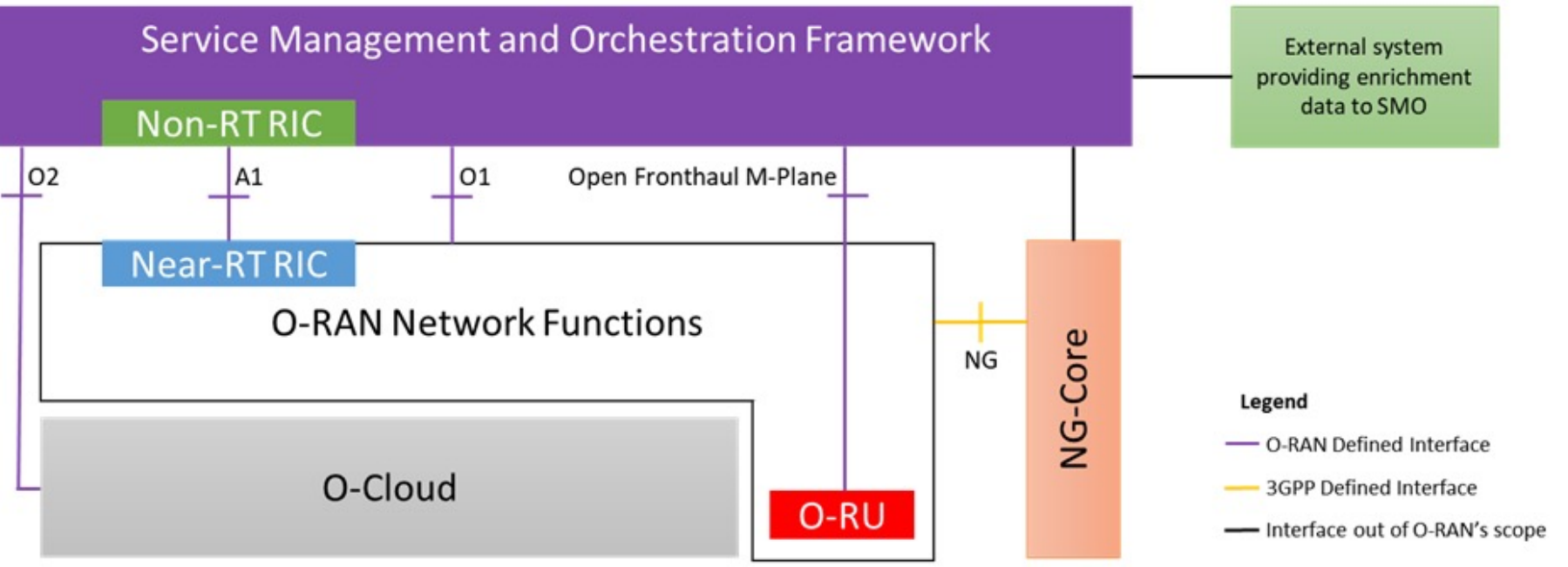}
    \caption{O-RAN's high-level architecture}
    \label{fig:high_level_arch}
\end{figure}

    \noindent Interface to O-RAN Network Functions; the management categories for which the international organization for standardization (ISO) model defines network management tasks (FCAPS: Fault, Configuration, Accounting, Performance, Security).
    \noindent Non-RT RIC for optimizing the RAN resources 
O-Cloud Management, Orchestration, and Workflow Management.

The Open Systems Interconnection model (OSI model), which standardizes and characterizes a telecommunication system communication functions, introduces seven abstraction layers ranging from the physical communications medium and up to the application layer (layers: Physical, Datalink, Network, Transport, Session, Presentation, Application). 3GPP specifications describe the NR radio interface protocol stack (a group of protocols that work together to provide infrastructure for networking access and activities) as layer 1, layer 2 and layer 3 protocols (or L1, L2 and L3 respectively), corresponding to the OSI model layers as follows: 1) Physical, 2) Data link and 3) Network. Data flows between the stack layers in channels. Transport channels (between layer 1 and layer 2) are responsible for the physical transfer of the data, and logical channels (between layer 2 and layer 3) relate to the type of data being transferred. 

The high-level architecture depicts four principle interfaces (A1, O1, Open Fronthaul M-plane and O2), which are used for connecting the SMO framework to O-RAN network functions and O-Cloud.

\noindent \textbf{O1 Interface:} responsible for FCAPS support of the O-RAN network functions which are O-Cloud hosted VNFs (virtualized or containerized network functions) and/or PNFs (physical network functions) running on customized hardware. 

\noindent \textbf{O2 Interface:} responsible for workload and resource management of the O-Cloud, a cloud computing platform consisting of physical infrastructure fulfilling O-RAN’s requirements. O-RU stands for O-RAN compliant RU, which is a logical node that hosts the Low-PHY layer and RF processing based on a lower layer functional split. 
    
\noindent \textbf{Open Fronthaul M-plane (Management Plane) Interface:} responsible for O-RU administration activities; architecturally it is used to support O-RU management in a hybrid model, where O-RU is managed by one or more NMSs (Network Management Systems), in addition to O-DUs (O-RAN Distributed Units), as opposed to the hierarchical model in which the O-RU is managed solely by O-DUs. Standardizing the O-RU management functions aligns with O-RAN’s multi-vendor RAN goal, eliminating the dependency on specific vendors. 
    
\noindent \textbf{A1 Interface:} responsible for connecting the Near-RT RIC and Non-RT RIC.

Figure 4 depicts the 5G RAN disaggregated architecture, which splits the RAN into functional units: the RIC (RAN Intelligent Controller), centralized unit (CU), distributed unit (DU), and RU. The DU and CU inherited the base station’s digital processing computation roles, introducing digital data into the network, and different workloads are split among the DU and CU based on their respective latency requirements. 
At a high level, the RIC \cite{balasubramanian2021ric} is a software-defined network (SDN) based component that performs selected radio resource management (RRM) functions that control the network’s operation. In O-RAN’s architecture, the RIC hosts the machine learning capabilities that drive the radio network’s automation capability. Figure 3 depicts two RIC components: a Non-Real-Time RIC (Non-RT RIC) and a Near Real-Time RIC (Near-RT RIC), which are connected by an A1 interface. In O-RAN’s architecture, the Non-RT RIC resides within the SMO, and typically the execution time of its control loop is one second or more, while the Near-RT RIC control loops run in the order of 10 milliseconds or more. 

\begin{figure}
    \centering
    \includegraphics[width=0.5 \textwidth]{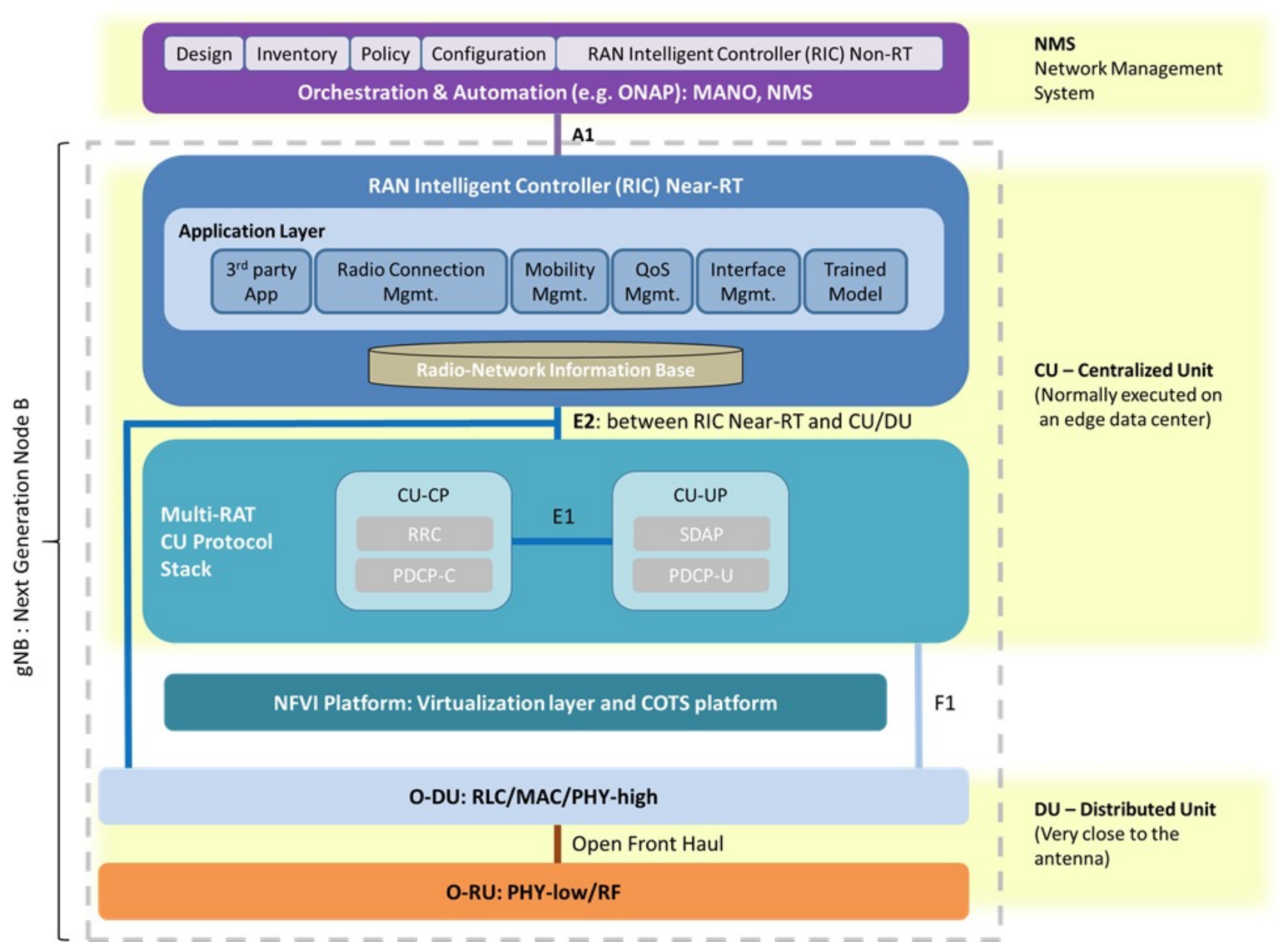}
    \caption{O-RAN’s reference architecture, logical entities, and interfaces \cite{alliance2020ran}: the disaggregated radio access network (RAN) introduces new technologies and therefore, new threats to the cellular network }
    \label{fig:oran_reference_arch}
\end{figure}

    \noindent \textbf{Non-RT RIC:} its main goal is to support intelligent RAN optimization by using guidance policies, machine learning model management, and information enrichment for the Near-RT RIC. It is also responsible for running modular applications (rApp via its R1 interface, not shown in the diagram) which uses the Non-RT RIC framework’s capabilities to perform RAN optimization as well as other tasks through its interfaces. 
    
    \noindent \textbf{Near-RT RIC:} a logical function that enables near real-time control and optimization of RAN elements and resources utilizing machine learning models. 
    
    \noindent \textbf{Near-RT RIC Applications:} Near-RT RIC can host xApp applications which can be provided by different third parties and can serve as the baseline resource management use cases or support new and unknown use cases.
    
    \noindent \textbf{Centralized Unit (CU):} CU is responsible for non-real-time and higher L2, and L3. L1, L2 and L3 are OSI model-based split of duties in the protocols required to enable the distribution of the RAN functionality. 
    
    \noindent \textbf{Control User Plane Separation (CUPS):} The CU can be further divided into its control plane (CP) and user plane (UP) functions which improve the placement of different RAN functions, thereby accommodating different situations and performance needs (the CU-CP and CU-UP, respectively). The CU-CP and CU-UP interface is known as E1, and it is a CP interface. The O-CU-CP (O-RAN centralized unit control plane) is a logical node hosting the radio resource control (RRC) and the CP part of the packet data convergence protocol (PDCP). The O CU-UP (O-RAN centralized unit user plane) is a logical node hosting the service data adaptation protocol (SDAP) and the user plane (UP) part of PDCP protocol. 
    
    \noindent \textbf{Multi-RAT CU Protocol Stack:} Supports various protocol stacks, for example, 4G and 5G multiple radio access technologies in the same network, and is also relevant for split option 7.2x for use cases where efficient resource utilization from multi-RAT may be needed. 
    
    \noindent \textbf{Distributed Unit (DU):} The DU is responsible for real-time L1 and L2 scheduling functions. 
    
    \noindent \textbf{CU and DU split:} There are several CU and DU split options tailored to use case, scenario, and performance/bandwidth trade-offs \cite{chaudhary2019c}, however the most common definition of the O-RAN Alliance is Option 7.2, where the protocol stack is split and the O-RU hosts the Low-PHY (physical layer) and the RF parts, the O-DU hosts the Hi-PHY, medium access control (MAC), and radio link control (RLC), and the O-CU hosts the packet data convergence protocol (PDCP), service data adaptation protocol (SDAP) and the radio resource control (RRC). 
    
    \noindent \textbf{Radio Unit (RU):} The RU is responsible for radio frequency signals broadcast and transmission, and it is usually part of the antenna.

The RAN and Mobile network have intermediate communication links between their components (see Figure 2).

    \noindent \textbf{Fronthaul:} Connects the RRH and RU to the digital processing equipment. O-RAN has defined an Open Front Haul interface between the O-RU and O-DU, breaking the single-vendor proprietary implementations and allowing different players to provide radio and distributed units on top of a standardized interface. 
    
    \noindent \textbf{Midhaul:} Known also as the F1 interface, is the communication link between the O-DU and O-CU. 
    
    \noindent \textbf{Backhaul:} This is used to connect the RAN towards the core network.

\section{\label{sec:ontology} Security Evaluation Ontology}
\subsection{Overview}
One of the main challenges in evaluating the security of O-RAN is the fact it is currently at an early development stage. Over the past two years, we have witnessed extensive efforts invested by the O-RAN Alliance in promoting this development. However, in practice, the current development release (i.e., the Dawn version) is far from final or stable, and the technology stack used in O-RAN changes frequently. In addition, sometimes, the technology stack differs among the various consortiums promoting the O-RAN concept.
Existing methodologies for evaluating enterprise security risk (such as the methodology presented by NIST \cite{singhal2012security}) require a high level of details of the target mostly based on an advanced level of implementation maturity. As such, they cannot be used without changes for assessing the security risk of a system based only on its high-level architecture and design documents. For instance, the ontology presented by NIST (widely adopted by security practitioners) considers entities such as the specific hardware and software used by the system. Such entities, for example, are not described in O-RAN’s specification.

To address the above-mentioned challenge, we devised a specialized ontology to enable the security risk assessment of a system that does not include specific implementation details, where the ontology is inspired by NIST framework. The security evaluation process initial step is the definition of the ontology whereas the next step is a definition of a detailed taxonomy covering all the entities and activities in the system, relevant from a security analysis point of view. Following the taxonomy definition, the concrete threat survey methodology is described explaining how we identified the threats in the literature, and the final step is categorizing the identified threats into the taxonomy. The primary benefits of such an early-stage security risk assessment are threefold: First, it enables the identification of potential security threats during the system design phase, when design changes are easier to implement. Second, it enables the early identification and implementation of security countermeasures to mitigate these threats as well as architectural changes which can enable usage of such countermeasures. Third, it creates a baseline risk management framework that can evolve alongside O-RAN evolution and serve operators in their overall risk strategy.

\subsection{The Proposed Ontology}
\begin{figure}
    \centering
    \includegraphics[width=0.5\textwidth]{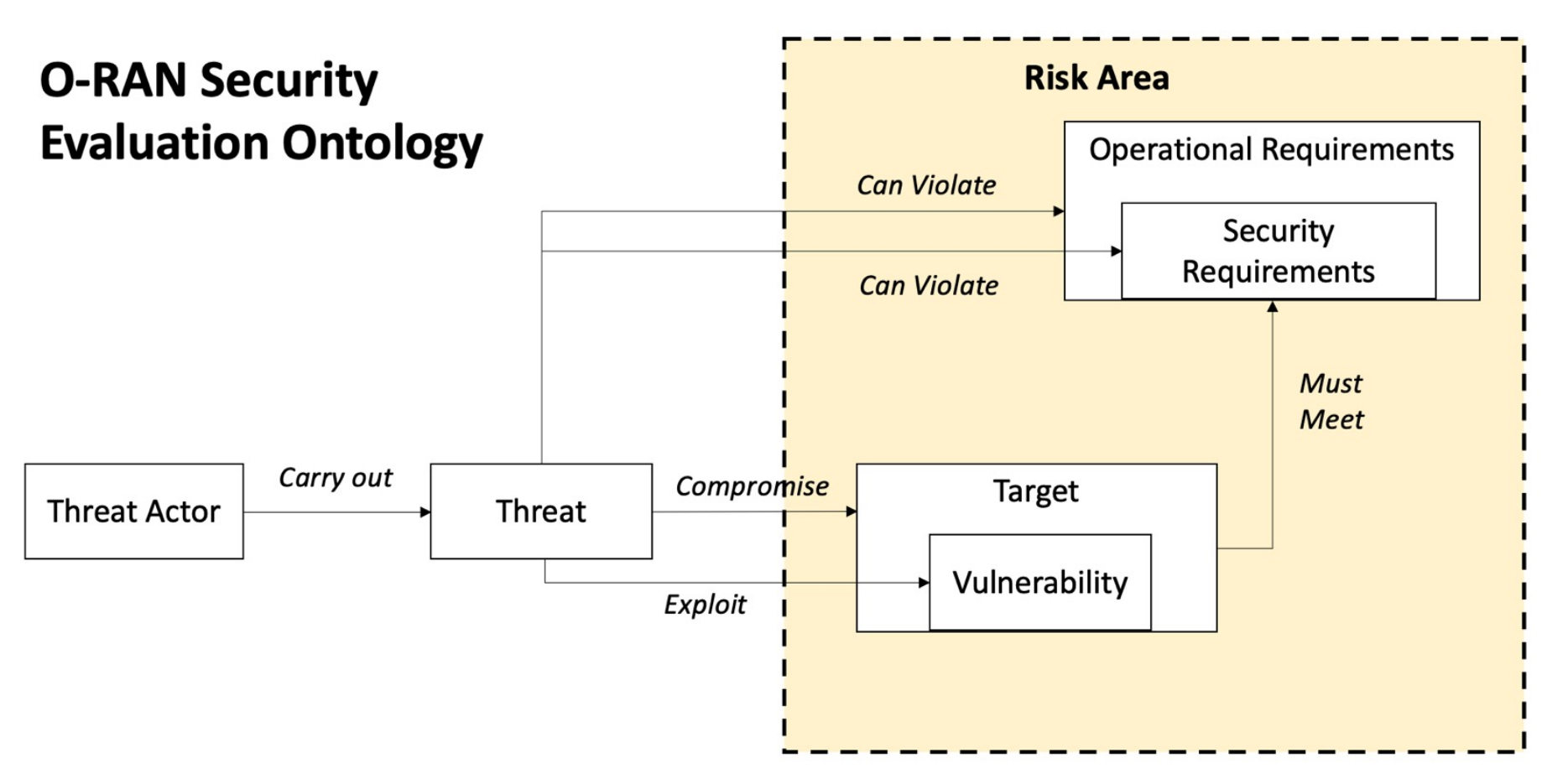}
    \caption{The ontology we devised for the security analysis.}
    \label{fig:ontology}
\end{figure}

Concretely, we propose the following ontology for the security evaluation of O-RAN where it includes the following seven entities (see Figure~\ref{fig:ontology}): Threat Actor, Threat, Risk Area, Target, Vulnerability, Security Requirements, Operational Requirements. In this section, we briefly describe those entities and the relationships between them. 

\noindent\textbf{Risk Area:} A semantic grouping of different threats based on a specific major aspect of O-RAN architecture.
    
\noindent\textbf{Threat Actor:} An individual, group, or state carrying out a threat or range of threats against a system. 
    
\noindent\textbf{Threat:} Concrete set of actions aimed at compromising a target which can result in a potential violation of security and/or operational requirements of a given target within a given system.
    
\noindent\textbf{Target:} A target is a component (i.e. technology, hardware element, software element…) within the system that must meet the operational and security requirements of the system and is at risk of compromise due to a certain threat. Important to note that once a target is compromised by a threat, optionally an additional set of related components can become targets for lateral risk though for sake of brevity and due to the lack of implementation details in the O-RAN spec we have not covered that extended aspect.
    
\noindent\textbf{Vulnerability:} A flaw in a target component that can be exploited by a threat utilizing one or more attack techniques. Important to note that due to the lack of implementation details in the O-RAN spec it is impossible to detail attack techniques and are omitted from this evaluation.
    
\noindent\textbf{Operational and Security Requirements:} The operational requirements that each component within a system should comply with to function properly. The security requirements are part of the overall operational requirements of a system. A threat that was carried out successfully can impact the system’s ability to function or comply with its security requirements. 

\subsection{Threat Analysis Methodology}
We define the following methodology for reviewing past threat cases and validating their relevance to O-RAN.
The methodology involves the following two main phases described below. 
\begin{enumerate}
    \item \textbf{Iterate over list of risk areas:}
    \begin{enumerate}
        \item For each respective risk area we enumerate the technologies, functions, environments, frameworks, and concepts used in O-RAN.
        
        \item Discover a list of threats known in academic papers as well as in industry publications that are relevant to the identified technologies, functions, environments, frameworks, and concepts from the previous step.
        
        \item For each threat:
            \begin{enumerate}
                \item Collect information on the characteristics of the actor.
                
                \item Identify the original target of attack. 
                
                \item Understand the vulnerability that resides in the targets.
                
                \item Identify the threat model that can exploit the vulnerability including attack techniques. Exclude alternative threat models and alternative attack techniques for the sake of brevity.
                
                \item Identify the security and overall operational impact of the identified threat.
                
                \item Exclude threats from the list which are not relevant
            \end{enumerate}
    \end{enumerate}

    \item \textbf{Enumerate the list of identified relevant threats from the previous step where for each threat project the threat attributes into the world of O-RAN in the respective risk area:}
    
    \begin{enumerate}
        \item Map potential targets within O-RAN that are seemingly susceptible to the threat based on the identified threat model and vulnerabilities.
        
        \item Identify the vulnerabilities mapping into the world of O-RAN

        \item Identify a list of relevant actors in the world of O-RAN which can carry out the threat.
        
        \item Deduce the threat operational range requirements.
        
        \item Evaluate threat transferability considering the O-RAN architecture.
        
        \item Identify potentially impacted security and operational requirements. 
    \end{enumerate}
\end{enumerate}

\subsection{Taxonomy for Cybersecurity Threat Analysis in O-RAN}
\begin{figure*}[t]
    \centering
    \includegraphics[width=1\textwidth]{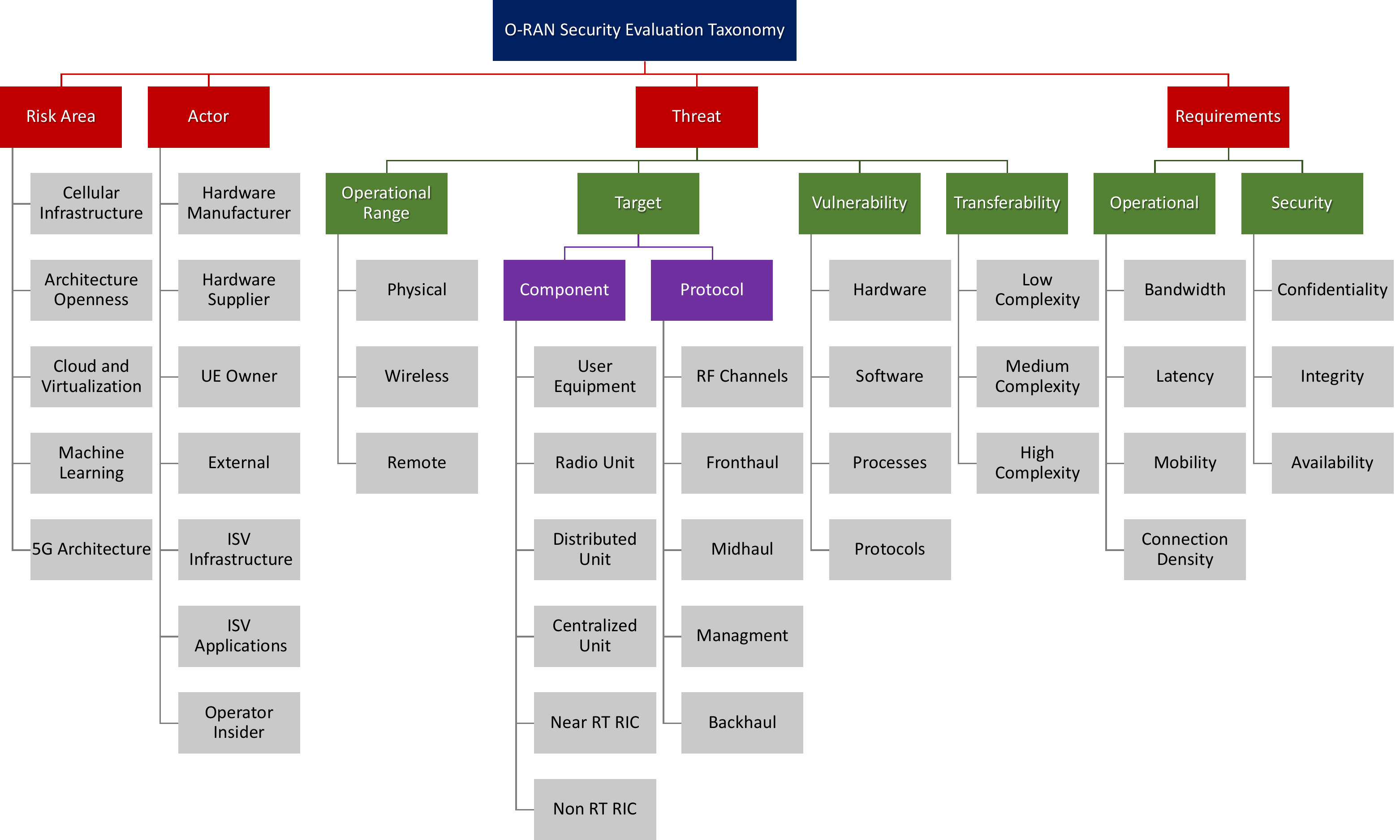}
    \caption{The taxonomy we devised for the security analysis of O-RAN.}
    \label{fig:taxonomy}
\end{figure*}

Based on the proposed ontology and methodology we conducted a comprehensive security analysis of O-RAN. A detailed description of the analysis in provided in Section \ref{sec:threat_analysis}.
In Figure~\ref{fig:taxonomy}), we present the product of this assessment: A Taxonomy for Cybersecurity Threat Analysis in O-RAN. The taxonomy map the different ontology entities to their practical instances within O-RAN. The goal of the taxonomy is to serve as a tool for assessing the overall attack surface of O-RAN. 

\section{Security Analysis}
In this section we describe the main findings of the proposed security analysis. 
Specifically, we enumerate the relevant risk areas, identify meaningful threat actors,  detect components within O-RAN that can be targeted by threats and classify the operational range that is required for executing a given threat.

\subsection{Risk Area}
In the proposed threat model, we have identified five material risk areas which embody the innovations in O-RAN as well as other general aspects of the architecture. These risk areas are our baseline for grouping the threats we identified which are relevant to O-RAN. The following are the five risk areas:

\noindent\textbf{Cellular Infrastructure:} Cellular networks have always been a target for attackers and there is a multitude of attacks, whether theoretical or real-life examples, which were aimed at the core principles of the cellular architecture. 

\noindent\textbf{Architectural Openness:} The main characteristic of openness and disaggregation within O-RAN opens a new risk area which is new to radio access networks in general.

\noindent\textbf{Cloud and Virtualization:} Cloud and virtualization comprise of a set of technologies, practices and processes developed in the world of public and private clouds and across that technological evolution it has been a prime target for many innovative threats.

\noindent\textbf{Machine Learning:} The world of machine learning is pervasive in different technological domains and due to the high level of interest it enjoys the risk in that area has been elevated with highly advanced and innovative attacks in the past. This is a highly emerging topic as the underlying technology itself evolves at a rapid pace.

\noindent\textbf{5G Architecture:} The 5G architecture depicts certain components, capabilities and topologies and that poses a new risk area for new types of attacks taking advantage of the architecture. 

\subsection{Threat Actor}
In the proposed threat model, we classified adversaries based on their engagement patterns with a potential O-RAN system whether at the design stage or during the operational stage. Since O-RAN offers a new type of system it involves new different actors borrowed from other domains. We identified the following threat actors: 

\noindent\textbf{Hardware Manufacturer:} The manufacturer of a hardware component used within an O-RAN system. The adversarial capability of such a threat actor is the ability to replace benign hardware components with malicious ones. Such malicious activity can operate in a standalone manner or can be controlled via an external entity such as in the case of a backdoor. This threat actor represents the difficulty to solve hardware supply chain problems where it is impossible to identify malicious from benign hardware components. The hardware manufacturer and the hardware supplier play a major role in the main theme of O-RAN openness and disaggregation.

\noindent\textbf{Hardware Supplier:}The supplier of a hardware component used by O-RAN whether as an integrator of systems or as part of the distribution supply chain. Note, in this threat model, we do not assume that the hardware component is manufactured by the supplier. In terms of adversarial capabilities, a hardware supplier has physical access to the hardware components used by O-RAN and has the capabilities to replace benign hardware components with malicious ones. Furthermore, in the case of a systems integrator, the supplier can also manipulate the firmware and operating system stack. It should be noted that in terms of attacker capabilities, a major difference between hardware manufacturer and hardware supplier is rooted in the ability of a hardware manufacturer to deploy security countermeasures that will prevent the supplier from carrying out a threat. 

\noindent\textbf{UE Owner:} Within the world of cellular networks the UE, User Equipment, was traditionally considered a personal cellular phone, while in the world of 5G, UEs can be of different types as the new category of IoT enables the connectivity of many devices operating in the real world. That expansion of the UE category creates a vast new attack surface on cellular networks. The UE based threat actor has a consumption type of relationship with the radio access network and due to that it is the least privileged type of actor while at the same time the vast amount of UEs and the ability to attack them with mobile computer viruses to gain control turns them into a unique threat actor. UEs can become malicious whether by supply chain threats in their product assembly or via attacks disguised as legitimate applications or software updates.  

\noindent\textbf{External:} An external actor refers to an undefined entity that can carry out certain types of threats in the various risk areas. It is a basket-type of an actor which aggregates different unique types of threats that do not fall into the well-defined other categories of threat actors.

\noindent\textbf{ISV Infrastructure:} The cloudification of O-RAN and the openness dictates working with different vendors for building the system and ISV providing infrastructure software, whether proprietary or open-source, plays a big role in the buildup of O-RAN. The adversarial capability of an ISV infrastructure threat actor is the ability to deploy malicious logic in the software infrastructure. Within the category of infrastructure ISV, we include the providers of operating systems, orchestration frameworks, runtime environments, O-RAN system components, IT tools, etc. 

\noindent\textbf{ISV Applications:} As one of the goals of O-RAN is to enable extensibility of the radio access network with new functionalities required to support new use cases the role of ISV application developers is dominant. The adversarial capability of this threat actor is the ability to deploy malicious logic in third-party applications running on top of the O-RAN system. It is hard to predict what type of applications will be developed on O-RAN beyond the known use cases of smart factories, autonomous driving and others and due to that this category of actors can become the major one within time. The major difference between ISV applications and the ISV infrastructure categories is the fact that applications are supposed to be of less privileged nature within the system and has much lower access to critical services and infrastructure as the architecture depicts separation between the infrastructure which operates the system and the applications enabled on top of it.

\noindent\textbf{Operator Insider:} An adversary who is a privileged system-level user within an O-RAN deployment, whether an employee at the operator company or a contractor, which has programmatic access to the O-RAN system.  Such a rogue insider can operate at different stages in the O-RAN system lifecycle and is assumed to have access which inherently poses risk. The category of insider includes the case where a legitimate operator of the system has been compromised and the stolen credentials are used to get insider-level access.

\subsection{Threat}
The threats identified in this analysis are categorized with the following criteria: 

\noindent\textbf{Operational Range (denoted as OR):} What is required for the actor to carry out successfully the threat in terms of location relative to its target.
We define three operational range levels: 
\begin{itemize}
    \item \textit{Physical (denoted as P):} The actor is required to have direct physical access to the target to carry out the threat.

    \item \textit{Wireless (denoted as W):} The actor is required to be within the wireless range of the RAN to carry out the threat.
    
    \item\textit{Remote (denoted as R):} The actor can be located remotely and still be able to carry out the threat. 
\end{itemize}

\noindent\textbf{Target:} The target is a defined asset within the system and this category includes the logical components and protocols in the O-RAN architecture. A target can include one or more targets that can be compromised under a certain threat. Important to note that detailed threat analysis is required once a complete implementation of O-RAN with detailed system design is available. For sake of brevity, we have created a simple list of components and protocols derived from the O-RAN architecture and defined it as options in this taxonomy. The available options for target components and protocols are described in Table~\ref{tab:target_components}. 
\setlength\tabcolsep{4.2pt}
\begin{table}[t!]
\scriptsize
\centering
{\begin{tabular}{|c"m{0.12\textwidth}"m{0.25\textwidth}|}
\Xhline{3\arrayrulewidth}

\makecell{\textbf{Type}} &
\textbf{Asset} &
\textbf{Description} \\
\Xhline{3\arrayrulewidth}

\multirow{6}{*}{\textbf{Components}} 
& User Equipment (UE) & The UE although not part of the cellular network still plays an inseparable role as it connects and communicates with the cellular network. \\\cline{2-3}

& Radio Unit (RU) & The RU is responsible on radio communications with UEs and digital transfer of the communications into the DU. \\\cline{2-3}

& Distributed Unit (DU) & The DU is a cloudified compute unit mostly responsible on digital processing of the communications arriving from the RU but also capable of running workload with very low latency requirements. \\\cline{2-3}

& Centralized Unit (CU) & The CU is a cloudified compute unit responsible on aggregating communications with multiple DUs and serves as a general-purpose hosting area for applications (edge) with mid-level latency requirements. \\\cline{2-3}

& Near-RT RIC  & The Near-RT RIC is responsible on optimizing the network resources (CU, DU, RU) to comply with requirements and policies arriving from the Non-RT RIC. The Near-RT RIC is where the system-level machine learning inference takes place. \\\cline{2-3}

& Non-RT RIC  & The non-RT RIC also called the SMO is the area which on hand connects to the core network and on the other hand responsible on managing the network from a configuration point of view. Among other roles of the SMO are billing, information collection, ML pipeline tasks and others. \\\cline{2-3}
\Xhline{3\arrayrulewidth}

\multirow{5}{*}{\makecell{\textbf{Protocols} }} 
& RF Channels & The RF channels are the concrete radio communications used by the UE and RU. \\\cline{2-3}

& Fronthaul & The Fronthaul is set of protocols governing the communications between the RU and the DU. \\\cline{2-3}

& Midhaul & The Midhaul is a set of protocols governing the communications between the CU and the DU. \\\cline{2-3}

& Management & The Management is a set of protocols governing the communications between the Near-RT RIC, the RIC and the CU, DU and RU utilized for managing the network. \\\cline{2-3}

& Backhaul & The Backhaul is a set of protocols governing the communications between the RAN and the Core network. \\\cline{2-3}

\Xhline{3\arrayrulewidth}

\end{tabular}}
\caption{Target components and protocols}
\label{tab:target_components}
\end{table}

\noindent\textbf{Vulnerability:} - This criterion defines in which area of the target the vulnerability resides where it can one of the following options: 
\begin{itemize}
    \item \textit{Hardware:} The vulnerability resides inside a hardware component whether it is part of the general-purpose compute used in O-RAN or specialized hardware related to the radio aspect of the network. The hardware category includes firmware-related attacks even though they are software-defined.
    
    \item \textit{Software:} The vulnerability resides in a software component in the broad sense where it includes virtualization layers, operating systems, tools and applications.
    
    \item\textit{Processes:} The vulnerability exists in a process whether it is a human-driven process or a process controlled by a computer. In the case of computer-based processes, the vulnerability.
    
    \item\textit{Protocol:} The vulnerability exists in a computer-based communications protocol.
\end{itemize}

\noindent\textbf{Transferability (denoted as TR):} This criterion assesses the complexity needed to modify and transfer an existing threat from the domain it was identified into O-RAN. Our evaluation of the complexity is based on our understanding of the way the threat operates and the way O-RAN is designed while it is not based on a rigorous evaluation of the threat. We define three transferability complexity levels: 

\begin{itemize}
    \item \textit{Low Complexity:} The threat can be applied directly on O-RAN without any modification of the attack techniques and details. 
    
    \item \textit{Medium Complexity:} The threat can be applied on O-RAN, but it requires some modification to the attack configuration and packaging.
    
    \item\textit{High Complexity:} The threat cannot be applied on O-RAN without reengineering the threat thoroughly while in general, on the conceptual level, the threat seems to apply to O-RAN.
\end{itemize}

\subsection{Requirements}
\noindent\textbf{Operational Requirements:} The operational requirements of O-RAN: Low Latency (Denoted as LL), Bandwidth (Denoted as B), Connection Density (Denoted as CD), Energy Efficiency (Denoted as EE), and Mobility (Denoted as M) that can be violated by a threat via compromising a target that must meet those requirements are detailed in Section II-B.

\noindent\textbf{Security Requirements:}The security requirements we use for evaluating the threat impact are: 

\begin{itemize}
    \item \textit{Confidentiality Impact:} Any type of threat that discloses information to unauthorized entities. 
    
    \item \textit{Integrity Impact:} Any type of threat that modifies or allows the modification of information processed by the asset. 
    
    \item\textit{Availability Impact:} Any type of threat that prevents the asset from operating.
\end{itemize}
\section{\label{sec:threat_analysis}O-RAN Attack Surface}
In this section, we review the list of identified threats divided into the respective risk areas where for each threat identified we map the relevant criteria as defined in the taxonomy. The longer list of non-relevant threats has been omitted for the sake of brevity. Furthermore, our approach is to list only a few research papers that focus on the same threat as there are threats that are covered in many research papers in a similar fashion. 

\setlength\tabcolsep{4.2pt}
\begin{table*}[tb]
\scriptsize
\centering
\hspace*{-1.65cm}
\begin{tabular}{
|m{0.205\textwidth}|  
m{0.23\textwidth}|  
m{0.004\textwidth}m{0.004\textwidth}m{0.004\textwidth}m{0.004\textwidth}
m{0.004\textwidth}m{0.004\textwidth}m{0.004\textwidth}m{0.004\textwidth}
m{0.004\textwidth}m{0.004\textwidth}m{0.004\textwidth}| 
m{0.004\textwidth}m{0.004\textwidth}m{0.004\textwidth}m{0.004\textwidth}| 
m{0.004\textwidth}m{0.004\textwidth}m{0.004\textwidth}m{0.004\textwidth}m{0.004\textwidth}
m{0.004\textwidth}m{0.004\textwidth}| 
m{0.03\textwidth}m{0.03\textwidth}| 
m{0.004\textwidth}m{0.004\textwidth}m{0.004\textwidth}m{0.004\textwidth}m{0.004\textwidth}| 
m{0.004\textwidth}m{0.004\textwidth}m{0.004\textwidth}|}  
\Xhline{3\arrayrulewidth}

\multirow{2}{*}{\makecell{\textbf{References}}} &
\multirow{2}{*}{\textbf{Threat}} &
\multicolumn{11}{c|}{\textbf{Asset}} & 
\multicolumn{4}{c|}{\textbf{Vuln.}} &
\multicolumn{7}{c|}{\textbf{Actor}} & 
\multicolumn{2}{c|}{\textbf{Mapping}} &
\multicolumn{5}{c|}{\textbf{\makecell{Operational\\Impact}}} &
\multicolumn{3}{c|}{\textbf{\makecell{Sec.\\Impact}}}
\\
\cline{3-34} 

& & 
\rotatebox{90}{User Equipment} & 
\rotatebox{90}{Radio Unit} &  
\rotatebox{90}{Distributed Unit} &
\rotatebox{90}{Centralized Unit} & 
\rotatebox{90}{Near-RT RIC} &
\rotatebox{90}{NNon-RT RIC} &
\rotatebox{90}{RF Channels} &
\rotatebox{90}{Fronthaul} &
\rotatebox{90}{Midhaul} &
\rotatebox{90}{Management} &
\rotatebox{90}{Backhaul} &

\rotatebox{90}{Protocol} &
\rotatebox{90}{Software} &
\rotatebox{90}{Hardware} &
\rotatebox{90}{Process} &

\rotatebox{90}{Hardware Manufacturer} &
\rotatebox{90}{Hardware Supplier} &
\rotatebox{90}{UE Owner} &
\rotatebox{90}{External} &
\rotatebox{90}{ISV Infrastructure} &
\rotatebox{90}{ISV Applications} &
\rotatebox{90}{Operator Insider} &

\rotatebox{90}{Operational Range} &
\rotatebox{90}{Transferability} &

\rotatebox{90}{Low Latency} &
\rotatebox{90}{Bandwidth} &
\rotatebox{90}{Connection Density} &
\rotatebox{90}{Energy Efficiency} &
\rotatebox{90}{Mobility} &

\rotatebox{90}{Confidentiality} &
\rotatebox{90}{Integrity}&
\rotatebox{90}{Availability} \\
\Xhline{3\arrayrulewidth}

\rowcolor{Grey}
\multicolumn{34}{|c|}{\textbf{Risk Area: Cellular Infrastructure}}\\ 
\cite{hussain20195greasoner,jover2013security,dehnel2018security,ferrag2018security,rupprecht2019breaking} & Cellular Protocol Vulnerabilities &
$\circ$ & 
$\bullet$ & 
$\circ$ & 
$\circ$ & 
$\circ$ & 
$\circ$ & 
$\bullet$ & 
$\circ$ & 
$\circ$ & 
$\circ$ & 
$\circ$ &  

$\bullet$ & 
$\circ$ & 
$\circ$ & 
$\circ$ & 
 
$\circ$ & 
$\circ$ & 
$\circ$ & 
$\bullet$ & 
$\circ$ & 
$\circ$ & 
$\circ$ & 

W & 
M & 

$\bullet$ & 
$\bullet$ & 
$\bullet$ & 
$\bullet$ & 
$\bullet$ &  

$\bullet$ & 
$\bullet$ & 
$\circ$ 
\\\hline 

\cite{kapetanovic2015physical} & Passive Eaves-dropping &
$\circ$ & 
$\circ$ & 
$\circ$ & 
$\circ$ & 
$\circ$ & 
$\circ$ & 
$\bullet$ & 
$\circ$ & 
$\circ$ & 
$\circ$ & 
$\circ$ &  

$\bullet$ & 
$\circ$ & 
$\circ$ & 
$\circ$ & 
 
$\circ$ & 
$\circ$ & 
$\circ$ & 
$\bullet$ & 
$\circ$ & 
$\circ$ & 
$\circ$ & 

W & 
H & 

$\circ$ & 
$\circ$ & 
$\circ$ & 
$\circ$ & 
$\circ$ &  

$\bullet$ & 
$\circ$ & 
$\circ$ 
\\\hline

\cite{kapetanovic2015physical,sodagari2015singularity,clancy2011efficient,miller2011vulnerabilities,vadlamani2016jamming,wang2009cooperative,jorswieck2005optimal,brady2006spatially,yang2007joint,farahmand2008anti,chi2008effects,mehdi2009analysis,mukherjee2010equilibrium,asadullah2009joint} & Jamming Attacks &
$\circ$ & 
$\bullet$ & 
$\circ$ & 
$\circ$ & 
$\circ$ & 
$\circ$ & 
$\bullet$ & 
$\circ$ & 
$\circ$ & 
$\circ$ & 
$\circ$ &  

$\bullet$ & 
$\circ$ & 
$\bullet$ & 
$\circ$ & 
 
$\circ$ & 
$\circ$ & 
$\circ$ & 
$\bullet$ & 
$\circ$ & 
$\circ$ & 
$\circ$ & 

W & 
H & 

$\bullet$ & 
$\bullet$ & 
$\bullet$ & 
$\circ$ & 
$\bullet$ &  

$\circ$ & 
$\circ$ & 
$\bullet$ 
\\\hline 

\cite{hong2018guti,hussain2019privacy,rupprecht2020imp4gt} & Side Channels  &
$\circ$ & 
$\circ$ & 
$\circ$ & 
$\circ$ & 
$\circ$ & 
$\circ$ & 
$\bullet$ & 
$\circ$ & 
$\circ$ & 
$\circ$ & 
$\circ$ &  

$\bullet$ & 
$\circ$ & 
$\circ$ & 
$\circ$ & 
 
$\circ$ & 
$\circ$ & 
$\circ$ & 
$\bullet$ & 
$\circ$ & 
$\circ$ & 
$\circ$ & 

W & 
H & 

$\circ$ & 
$\circ$ & 
$\circ$ & 
$\circ$ & 
$\circ$ &  

$\bullet$ & 
$\bullet$ & 
$\circ$ 
\\\hline 

\cite{acar2016sok,zhou2012dissecting,becher2011mobile,kotzias2020did,li2007study,alrawi2021circle} & UE Vulnerabilities &
$\bullet$ & 
$\circ$ & 
$\circ$ & 
$\circ$ & 
$\circ$ & 
$\circ$ & 
$\circ$ & 
$\circ$ & 
$\circ$ & 
$\circ$ & 
$\circ$ &  

$\circ$ & 
$\bullet$ & 
$\circ$ & 
$\bullet$ & 
 
$\circ$ & 
$\circ$ & 
$\circ$ & 
$\bullet$ & 
$\circ$ & 
$\circ$ & 
$\circ$ & 

R & 
H & 

$\circ$ & 
$\circ$ & 
$\circ$ & 
$\bullet$ & 
$\circ$ &  

$\circ$ & 
$\bullet$ & 
$\circ$ 
\\\hline 

\cite{hassija2019survey,shakhov2017energy,spreitzer2017systematic} & Application Attacks &
$\bullet$ & 
$\bullet$ & 
$\circ$ & 
$\circ$ & 
$\circ$ & 
$\circ$ & 
$\circ$ & 
$\circ$ & 
$\circ$ & 
$\circ$ & 
$\circ$ &  

$\bullet$ & 
$\circ$ & 
$\circ$ & 
$\circ$ & 
 
$\circ$ & 
$\circ$ & 
$\bullet$ & 
$\bullet$ & 
$\circ$ & 
$\circ$ & 
$\circ$ & 

W & 
M & 

$\circ$ & 
$\circ$ & 
$\circ$ & 
$\bullet$ & 
$\circ$ &  

$\circ$ & 
$\bullet$ & 
$\circ$ 
\\\hline 

\cite{spreitzer2017systematic} & Side Channel Attacks &
$\bullet$ & 
$\circ$ & 
$\circ$ & 
$\circ$ & 
$\circ$ & 
$\circ$ & 
$\circ$ & 
$\circ$ & 
$\circ$ & 
$\circ$ & 
$\circ$ &  

$\circ$ & 
$\circ$ & 
$\circ$ & 
$\circ$ & 
 
$\circ$ & 
$\circ$ & 
$\circ$ & 
$\bullet$ & 
$\circ$ & 
$\circ$ & 
$\circ$ & 

P & 
H & 

$\circ$ & 
$\circ$ & 
$\circ$ & 
$\circ$ & 
$\circ$ &  

$\bullet$ & 
$\circ$ & 
$\circ$ 
\\\hline 

\cite{antonakakis2017understanding,kolias2017ddos,bertino2017botnets,zhang2014iot,marzano2018evolution,edwards2016hajime} & UE Botnets and Malware &
$\bullet$ & 
$\circ$ & 
$\circ$ & 
$\circ$ & 
$\circ$ & 
$\circ$ & 
$\circ$ & 
$\circ$ & 
$\circ$ & 
$\circ$ & 
$\circ$ &  

$\circ$ & 
$\bullet$ & 
$\circ$ & 
$\bullet$ & 
 
$\circ$ & 
$\circ$ & 
$\bullet$ & 
$\bullet$ & 
$\circ$ & 
$\circ$ & 
$\circ$ & 

R & 
H & 

$\odot$ & 
$\odot$ & 
$\odot$ & 
$\odot$ & 
$\odot$ &  

$\circ$ & 
$\bullet$ & 
$\bullet$ 
\\\hline 

\cite{enck2005exploiting,racic2008exploiting,traynor2007attack,kim2015breaking,tu2015voice,yu2019effects} & DoS on Cellular Services & 
$\bullet$ & 
$\circ$ & 
$\circ$ & 
$\circ$ & 
$\circ$ & 
$\circ$ & 
$\circ$ & 
$\bullet$ & 
$\bullet$ & 
$\circ$ & 
$\bullet$ &  

$\bullet$ & 
$\bullet$ & 
$\circ$ & 
$\circ$ & 
 
$\circ$ & 
$\circ$ & 
$\circ$ & 
$\bullet$ & 
$\circ$ & 
$\circ$ & 
$\circ$ & 

R & 
M & 

$\bullet$ & 
$\bullet$ & 
$\bullet$ & 
$\bullet$ & 
$\bullet$ &  

$\circ$ & 
$\circ$ & 
$\bullet$ 
\\\hline 

\cite{yan2015software} & DDoS flooding attacks &
$\bullet$ & 
$\circ$ & 
$\circ$ & 
$\circ$ & 
$\circ$ & 
$\circ$ & 
$\circ$ & 
$\bullet$ & 
$\bullet$ & 
$\circ$ & 
$\bullet$ &  

$\bullet$ & 
$\bullet$ & 
$\circ$ & 
$\circ$ & 
 
$\circ$ & 
$\circ$ & 
$\circ$ & 
$\bullet$ & 
$\circ$ & 
$\circ$ & 
$\circ$ & 

R & 
M & 

$\bullet$ & 
$\bullet$ & 
$\bullet$ & 
$\bullet$ & 
$\bullet$ &  

$\circ$ & 
$\circ$ & 
$\bullet$ 
\\\hline 

\cite{shin2013attacking} & DoS on control plane &
$\circ$ & 
$\circ$ & 
$\bullet$ & 
$\bullet$ & 
$\bullet$ & 
$\bullet$ & 
$\circ$ & 
$\circ$ & 
$\bullet$ & 
$\circ$ & 
$\bullet$ &  

$\bullet$ & 
$\bullet$ & 
$\circ$ & 
$\circ$ & 
 
$\circ$ & 
$\circ$ & 
$\circ$ & 
$\bullet$ & 
$\circ$ & 
$\circ$ & 
$\circ$ & 

R & 
M & 

$\bullet$ & 
$\bullet$ & 
$\bullet$ & 
$\bullet$ & 
$\bullet$ &  

$\circ$ & 
$\circ$ & 
$\bullet$ 
\\\hline 

\cite{rupprecht2016putting,hussain2018lteinspector,raza2017exposing,kim2019touching}& Vulnerability Detection Frameworks &
$\bullet$ & 
$\bullet$ & 
$\bullet$ & 
$\bullet$ & 
$\bullet$ & 
$\bullet$ & 
$\bullet$ & 
$\bullet$ & 
$\bullet$ & 
$\bullet$ & 
$\bullet$ &  

$\bullet$ & 
$\bullet$ & 
$\bullet$ & 
$\bullet$ & 
 
$\circ$ & 
$\circ$ & 
$\circ$ & 
$\bullet$ & 
$\circ$ & 
$\circ$ & 
$\circ$ & 

R,W & 
M,L & 

$\circ$ & 
$\circ$ & 
$\circ$ & 
$\circ$ & 
$\circ$ &  

$\circ$ & 
$\circ$ & 
$\circ$ 
\\

\Xhline{3\arrayrulewidth}
\rowcolor{Grey}
\multicolumn{34}{|c|}{\textbf{Risk Area: Architectural Openness}}\\ 
\cite{traynor2007attack}& Human Errors and Misconfiguration &
$\bullet$ & 
$\bullet$ & 
$\bullet$ & 
$\bullet$ & 
$\bullet$ & 
$\bullet$ & 
$\bullet$ & 
$\bullet$ & 
$\bullet$ & 
$\bullet$ & 
$\bullet$ &  

$\bullet$ & 
$\bullet$ & 
$\bullet$ & 
$\bullet$ & 
 
$\bullet$ & 
$\bullet$ & 
$\bullet$ & 
$\circ$ & 
$\bullet$ & 
$\bullet$ & 
$\bullet$ & 

P,W,R & 
H & 

$\bullet$ & 
$\bullet$ & 
$\bullet$ & 
$\bullet$ & 
$\bullet$ &  

$\bullet$ & 
$\bullet$ & 
$\bullet$ 
\\\hline 

\cite{sasaki2020security,nist-800-161} & Supply Chain - Software & 
$\bullet$ & 
$\bullet$ & 
$\bullet$ & 
$\bullet$ & 
$\bullet$ & 
$\bullet$ & 
$\bullet$ & 
$\bullet$ & 
$\bullet$ & 
$\bullet$ & 
$\bullet$ &  

$\bullet$ & 
$\bullet$ & 
$\circ$ & 
$\bullet$ & 
 
$\circ$ & 
$\circ$ & 
$\circ$ & 
$\circ$ & 
$\bullet$ & 
$\bullet$ & 
$\bullet$ & 

R & 
H & 

$\bullet$ & 
$\bullet$ & 
$\bullet$ & 
$\bullet$ & 
$\bullet$ &  

$\bullet$ & 
$\bullet$ & 
$\bullet$ 
\\\hline 

\cite{sasaki2020security,nist-800-161} & Supply Chain - Hardware & 
$\bullet$ & 
$\bullet$ & 
$\bullet$ & 
$\bullet$ & 
$\bullet$ & 
$\bullet$ & 
$\bullet$ & 
$\bullet$ & 
$\bullet$ & 
$\bullet$ & 
$\bullet$ &  

$\circ$ & 
$\circ$ & 
$\bullet$ & 
$\circ$ & 
 
$\bullet$ & 
$\bullet$ & 
$\bullet$ & 
$\circ$ & 
$\circ$ & 
$\circ$ & 
$\circ$ & 

R & 
M & 

$\bullet$ & 
$\bullet$ & 
$\bullet$ & 
$\bullet$ & 
$\bullet$ &  

$\bullet$ & 
$\bullet$ & 
$\bullet$ 
\\\hline 

\cite{mary2015shellshock}& Vulnerable Open-Source Packages &
$\circ$ & 
$\circ$ & 
$\bullet$ & 
$\bullet$ & 
$\bullet$ & 
$\bullet$ & 
$\circ$ & 
$\circ$ & 
$\circ$ & 
$\circ$ & 
$\circ$ &  

$\circ$ & 
$\bullet$ & 
$\circ$ & 
$\circ$ & 
 
$\circ$ & 
$\circ$ & 
$\circ$ & 
$\bullet$ & 
$\circ$ & 
$\circ$ & 
$\bullet$ & 

R & 
H & 

$\bullet$ & 
$\bullet$ & 
$\bullet$ & 
$\bullet$ & 
$\bullet$ &  

$\bullet$ & 
$\bullet$ & 
$\bullet$ 
\\\hline 

\cite{zhang2018code} & API Exploitation &
$\circ$ & 
$\circ$ & 
$\bullet$ & 
$\bullet$ & 
$\bullet$ & 
$\bullet$ & 
$\circ$ & 
$\circ$ & 
$\circ$ & 
$\circ$ & 
$\circ$ &  

$\circ$ & 
$\bullet$ & 
$\circ$ & 
$\circ$ & 
 
$\circ$ & 
$\circ$ & 
$\circ$ & 
$\bullet$ & 
$\bullet$ & 
$\bullet$ & 
$\bullet$ & 

R & 
L & 

$\bullet$ & 
$\bullet$ & 
$\bullet$ & 
$\bullet$ & 
$\bullet$ &  

$\bullet$ & 
$\bullet$ & 
$\bullet$ 
\\
\Xhline{3\arrayrulewidth}

\rowcolor{Grey}
\multicolumn{34}{|c|}{\textbf{Risk Area: Cloud and Virtualization}}\\ 

\cite{kocher2019spectre}& Co-hosted Application Side Channels &
$\circ$ & 
$\bullet$ & 
$\bullet$ & 
$\bullet$ & 
$\bullet$ & 
$\bullet$ & 
$\circ$ & 
$\circ$ & 
$\circ$ & 
$\circ$ & 
$\circ$ &  

$\circ$ & 
$\bullet$ & 
$\bullet$ & 
$\circ$ & 
 
$\circ$ & 
$\circ$ & 
$\circ$ & 
$\circ$ & 
$\bullet$ & 
$\bullet$ & 
$\bullet$ & 

R,P & 
H & 

$\bullet$ & 
$\bullet$ & 
$\bullet$ & 
$\bullet$ & 
$\bullet$ &  

$\bullet$ & 
$\bullet$ & 
$\bullet$ 
\\\hline  

\cite{almorsy2016analysis,pattaranantakul2019moving,shu2017study,zhang2012deep,ristenpart2009hey,lin2018measurement}& Image Manipulation &
$\circ$ & 
$\bullet$ & 
$\bullet$ & 
$\bullet$ & 
$\bullet$ & 
$\bullet$ & 
$\circ$ & 
$\circ$ & 
$\circ$ & 
$\circ$ & 
$\circ$ &  

$\circ$ & 
$\bullet$ & 
$\circ$ & 
$\circ$ & 
 
$\circ$ & 
$\circ$ & 
$\circ$ & 
$\circ$ & 
$\bullet$ & 
$\bullet$ & 
$\bullet$ & 

R,P & 
H & 

$\bullet$ & 
$\bullet$ & 
$\bullet$ & 
$\bullet$ & 
$\bullet$ &  

$\bullet$ & 
$\bullet$ & 
$\bullet$ 
\\\hline 

\cite{ristenpart2009hey,sgandurra2016evolution,zhang2012cross,zhang2014cross,liu2015last,Liyanage2017} & Guest-to-Guest Attacks &
$\circ$ & 
$\bullet$ & 
$\bullet$ & 
$\bullet$ & 
$\bullet$ & 
$\bullet$ & 
$\circ$ & 
$\circ$ & 
$\circ$ & 
$\circ$ & 
$\circ$ &  

$\circ$ & 
$\bullet$ & 
$\circ$ & 
$\circ$ & 
 
$\circ$ & 
$\circ$ & 
$\circ$ & 
$\circ$ & 
$\bullet$ & 
$\bullet$ & 
$\bullet$ & 

R,P & 
H & 

$\bullet$ & 
$\bullet$ & 
$\bullet$ & 
$\bullet$ & 
$\bullet$ &  

$\bullet$ & 
$\bullet$ & 
$\bullet$ 
\\\hline 

\cite{Liyanage2017,zhang2012deep,huang2012security,atya2017stalling}& Guest-to Hypervisor Attacks &
$\circ$ & 
$\bullet$ & 
$\bullet$ & 
$\bullet$ & 
$\bullet$ & 
$\bullet$ & 
$\circ$ & 
$\circ$ & 
$\circ$ & 
$\circ$ & 
$\circ$ &  

$\circ$ & 
$\bullet$ & 
$\circ$ & 
$\circ$ & 
 
$\circ$ & 
$\circ$ & 
$\circ$ & 
$\circ$ & 
$\bullet$ & 
$\bullet$ & 
$\bullet$ & 

R,P & 
H & 

$\bullet$ & 
$\bullet$ & 
$\bullet$ & 
$\bullet$ & 
$\bullet$ &  

$\bullet$ & 
$\bullet$ & 
$\bullet$ 
\\\hline 

\cite{ferrari2020nosql,continella2018there,Liyanage2017}& Inconsistent Security Policies &
$\circ$ & 
$\bullet$ & 
$\bullet$ & 
$\bullet$ & 
$\bullet$ & 
$\bullet$ & 
$\circ$ & 
$\circ$ & 
$\circ$ & 
$\circ$ & 
$\circ$ &  

$\bullet$ & 
$\bullet$ & 
$\circ$ & 
$\circ$ & 
 
$\circ$ & 
$\circ$ & 
$\circ$ & 
$\bullet$ & 
$\bullet$ & 
$\bullet$ & 
$\bullet$ & 

R,P & 
L & 

$\bullet$ & 
$\bullet$ & 
$\bullet$ & 
$\bullet$ & 
$\bullet$ &  

$\bullet$ & 
$\bullet$ & 
$\bullet$ 
\\\hline 

\cite{stuttard2011web,singh2016sql} & Exploiting Public-Facing Applications &
$\circ$ & 
$\bullet$ & 
$\bullet$ & 
$\bullet$ & 
$\bullet$ & 
$\bullet$ & 
$\circ$ & 
$\circ$ & 
$\circ$ & 
$\circ$ & 
$\circ$ &  

$\circ$ & 
$\bullet$ & 
$\circ$ & 
$\circ$ & 
 
$\circ$ & 
$\circ$ & 
$\circ$ & 
$\bullet$ & 
$\circ$ & 
$\circ$ & 
$\bullet$ & 

R,P & 
L & 

$\bullet$ & 
$\bullet$ & 
$\bullet$ & 
$\bullet$ & 
$\bullet$ &  

$\bullet$ & 
$\bullet$ & 
$\bullet$ 
\\\hline 

\cite{stuttard2011web} & Trusted Connected Third Parties &
$\circ$ & 
$\bullet$ & 
$\bullet$ & 
$\bullet$ & 
$\bullet$ & 
$\bullet$ & 
$\circ$ & 
$\circ$ & 
$\circ$ & 
$\circ$ & 
$\circ$ &  

$\circ$ & 
$\bullet$ & 
$\circ$ & 
$\circ$ & 
 
$\circ$ & 
$\circ$ & 
$\circ$ & 
$\bullet$ & 
$\bullet$ & 
$\bullet$ & 
$\bullet$ & 

R,P & 
M & 

$\bullet$ & 
$\bullet$ & 
$\bullet$ & 
$\bullet$ & 
$\bullet$ &  

$\bullet$ & 
$\bullet$ & 
$\bullet$ 
\\\hline 

\cite{he2015vetting,clark2013sok} & Improper SSL/TLS Configuration &
$\circ$ & 
$\bullet$ & 
$\bullet$ & 
$\bullet$ & 
$\bullet$ & 
$\bullet$ & 
$\circ$ & 
$\circ$ & 
$\circ$ & 
$\circ$ & 
$\circ$ &  

$\bullet$ & 
$\bullet$ & 
$\circ$ & 
$\circ$ & 
 
$\circ$ & 
$\circ$ & 
$\circ$ & 
$\bullet$ & 
$\bullet$ & 
$\bullet$ & 
$\bullet$ & 

R,P & 
H & 

$\bullet$ & 
$\bullet$ & 
$\bullet$ & 
$\bullet$ & 
$\bullet$ &  

$\bullet$ & 
$\bullet$ & 
$\bullet$ 
\\\hline 

\cite{patwary2020authentication} & Rogue Fog Nodes &
$\circ$ & 
$\bullet$ & 
$\bullet$ & 
$\bullet$ & 
$\bullet$ & 
$\bullet$ & 
$\circ$ & 
$\circ$ & 
$\circ$ & 
$\circ$ & 
$\circ$ &  

$\circ$ & 
$\bullet$ & 
$\circ$ & 
$\circ$ & 
 
$\circ$ & 
$\circ$ & 
$\circ$ & 
$\bullet$ & 
$\circ$ & 
$\circ$ & 
$\bullet$ & 

R,P & 
L & 

$\bullet$ & 
$\bullet$ & 
$\bullet$ & 
$\bullet$ & 
$\bullet$ &  

$\bullet$ & 
$\circ$ & 
$\circ$ 
\\
\Xhline{3\arrayrulewidth}

\rowcolor{Grey}
\multicolumn{34}{|c|}{\textbf{Risk Area: Machine Learning}}\\

\cite{he2019towards,wang2018stealing,suomalainen2020machine} & Misprediction  &
$\circ$ & 
$\circ$ & 
$\circ$ & 
$\circ$ & 
$\bullet$ & 
$\bullet$ & 
$\circ$ & 
$\circ$ & 
$\circ$ & 
$\circ$ & 
$\circ$ &  

$\bullet$ & 
$\bullet$ & 
$\circ$ & 
$\circ$ & 
 
$\circ$ & 
$\circ$ & 
$\circ$ & 
$\bullet$ & 
$\bullet$ & 
$\bullet$ & 
$\bullet$ & 

R,P & 
M & 

$\bullet$ & 
$\bullet$ & 
$\bullet$ & 
$\circ$ & 
$\bullet$ &  

$\circ$ & 
$\bullet$ & 
$\bullet$ 
\\\hline 

\cite{he2019towards,wang2018stealing,suomalainen2020machine} & Membership Attacks &
$\circ$ & 
$\circ$ & 
$\circ$ & 
$\circ$ & 
$\bullet$ & 
$\bullet$ & 
$\circ$ & 
$\circ$ & 
$\circ$ & 
$\circ$ & 
$\circ$ &  

$\bullet$ & 
$\bullet$ & 
$\circ$ & 
$\circ$ & 
 
$\circ$ & 
$\circ$ & 
$\circ$ & 
$\bullet$ & 
$\bullet$ & 
$\bullet$ & 
$\bullet$ & 

R,P & 
M & 

$\bullet$ & 
$\bullet$ & 
$\bullet$ & 
$\circ$ & 
$\bullet$ &  

$\bullet$ & 
$\circ$ & 
$\circ$ 
\\\hline 

\cite{he2019towards,wang2018stealing,suomalainen2020machine} & Training Data Extraction &
$\circ$ & 
$\circ$ & 
$\circ$ & 
$\circ$ & 
$\bullet$ & 
$\bullet$ & 
$\circ$ & 
$\circ$ & 
$\circ$ & 
$\circ$ & 
$\circ$ &  

$\bullet$ & 
$\bullet$ & 
$\circ$ & 
$\circ$ & 
 
$\circ$ & 
$\circ$ & 
$\circ$ & 
$\bullet$ & 
$\bullet$ & 
$\bullet$ & 
$\bullet$ & 

R,P & 
M & 

$\bullet$ & 
$\bullet$ & 
$\bullet$ & 
$\circ$ & 
$\bullet$ &  

$\bullet$ & 
$\bullet$ & 
$\circ$ 
\\\hline 

\cite{he2019towards,wang2018stealing,suomalainen2020machine} & Model Extraction &
$\circ$ & 
$\circ$ & 
$\circ$ & 
$\circ$ & 
$\bullet$ & 
$\bullet$ & 
$\circ$ & 
$\circ$ & 
$\circ$ & 
$\circ$ & 
$\circ$ &  

$\bullet$ & 
$\bullet$ & 
$\circ$ & 
$\circ$ & 
 
$\circ$ & 
$\circ$ & 
$\circ$ & 
$\bullet$ & 
$\bullet$ & 
$\bullet$ & 
$\bullet$ & 

R,P & 
M & 

$\bullet$ & 
$\bullet$ & 
$\bullet$ & 
$\circ$ & 
$\bullet$ &  

$\bullet$ & 
$\bullet$ & 
$\bullet$ 
\\\hline 

\rowcolor{Grey}
\multicolumn{34}{|c|}{\textbf{Risk Area: 5G Architecture}}\\

\cite{olimid20205g}& Slice Lifecycle Security &
$\circ$ & 
$\circ$ & 
$\circ$ & 
$\circ$ & 
$\bullet$ & 
$\bullet$ & 
$\circ$ & 
$\circ$ & 
$\circ$ & 
$\circ$ & 
$\circ$ &  

$\circ$ & 
$\bullet$ & 
$\circ$ & 
$\bullet$ & 
 
$\circ$ & 
$\circ$ & 
$\circ$ & 
$\bullet$ & 
$\bullet$ & 
$\bullet$ & 
$\bullet$ & 

R,P & 
M & 

$\bullet$ & 
$\bullet$ & 
$\bullet$ & 
$\circ$ & 
$\bullet$ &  

$\bullet$ & 
$\bullet$ & 
$\bullet$ 
\\\hline 

\cite{olimid20205g}& Intra-slice Security &
$\circ$ & 
$\bullet$ & 
$\bullet$ & 
$\bullet$ & 
$\circ$ & 
$\circ$ & 
$\circ$ & 
$\circ$ & 
$\circ$ & 
$\circ$ & 
$\circ$ &  

$\circ$ & 
$\bullet$ & 
$\circ$ & 
$\bullet$ & 
 
$\circ$ & 
$\circ$ & 
$\bullet$ & 
$\bullet$ & 
$\bullet$ & 
$\bullet$ & 
$\bullet$ & 

R,P & 
M & 

$\bullet$ & 
$\bullet$ & 
$\bullet$ & 
$\circ$ & 
$\bullet$ &  

$\bullet$ & 
$\bullet$ & 
$\bullet$ 
\\\hline 

\cite{olimid20205g} & Inter-slice Security & 
$\circ$ & 
$\circ$ & 
$\circ$ & 
$\circ$ & 
$\bullet$ & 
$\bullet$ & 
$\circ$ & 
$\circ$ & 
$\circ$ & 
$\circ$ & 
$\circ$ &  

$\circ$ & 
$\bullet$ & 
$\circ$ & 
$\bullet$ & 
 
$\circ$ & 
$\circ$ & 
$\circ$ & 
$\bullet$ & 
$\bullet$ & 
$\bullet$ & 
$\bullet$ & 

R,P & 
M & 

$\bullet$ & 
$\bullet$ & 
$\bullet$ & 
$\circ$ & 
$\bullet$ &  

$\bullet$ & 
$\bullet$ & 
$\bullet$ 
\\
\Xhline{3\arrayrulewidth}
\rowcolor{Blue}
\multicolumn{34}{|c|}{\textbf{Key Map}}\\
\multicolumn{34}{|l|}{Asset -  \textbf{$\bullet$}: The asset is being targeted by the threat, \textbf{$\circ$}: The asset is not being targeted by the threat}\\

\multicolumn{34}{|l|}{Vulnerability -  \textbf{$\bullet$}: The vulnerability resides within this type of component, \textbf{$\circ$}: The vulnerability does not resides within this type of component} \\

\multicolumn{34}{|l|}{Actor -  \textbf{$\bullet$}: The threat actor can obtain the capabilities required to materialize the threat ,\textbf{$\circ$}:The threat actor cannot obtain the capabilities required to materialize the attack.} \\

\multicolumn{34}{|l|}{Operational Range: \textbf{P} - \textbf{P}hysical, W - \textbf{W}ireless, \textbf{R} - \textbf{R}emote.}  \\

\multicolumn{34}{|l|}{Transferability: \textbf{L} - \textbf{L}ow, M - \textbf{M}edium, \textbf{H} - \textbf{H}igh.}  \\

\multicolumn{34}{|l|}{Operational Impact -  \textbf{$\bullet$}: The threat violates the operational requirement ,\textbf{$\circ$}:The threat does not violates the operational requirement.} \\

\multicolumn{34}{|l|}{Security Impact -  \textbf{$\bullet$}: The threat violates the security requirement ,\textbf{$\circ$}: The threat does not violates the security requirement.} \\
\Xhline{3\arrayrulewidth}

\end{tabular}
\caption{O-RAN Threat Mapping}
\label{tab:attack_techniques_with_actors}
\end{table*}

\subsection{Cellular Threats}
In this section, we focus on threats related to the risk area of cellular threats. We reviewed past threats that apply to RAN and cellular architectures and evaluated their applicability to O-RAN. We have specifically focused on threats related to the following assets: the RF channel, and the UE. Our review is based on three survey papers, which analyzed the security of traditional RAN \cite{tian2017survey,zou2016survey,mavoungou2016survey} while for each threat identified we have listed additional survey results. 

\subsubsection{The RF Channel}
Although, cellular networks have evolved dramatically in the past years. The communication channel between the user equipment and the remote radio head unit has not changed dramatically. That is, both traditional RAN and O-RAN use very similar RF channels. As a result, some of the known attacks on traditional RAN architectures theoretically can be easily transferred to O-RAN and we evaluated this risk. In our threat model, exploiting RF threats requires the actor to own a radio transceiver and to reside within the wireless range of the UE/RU. Thus, these threats can only be exploited by actors which are in the wireless range.  

\subsubsection{The UE}
The user’s equipment is a streamlined and legitimate access point to the cellular network where it enjoys certain privileges based on its subscription plan. As a result, a large number of attack case studies have been found that are focused on exploiting vulnerabilities in the user equipment and after successful compromise they turn into a threat for the cellular network. Threats on UEs are not bound to the architecture used by the cellular network in the first stage of compromising the UE equipment and therefore, these attacks apply to the O-RAN world. Within this survey we have listed threats that are aimed at compromising the UE equipment as well as threats that are posed from compromised UEs onto the cellular network. A major change between O-RAN and previous architectures is the increased diversity and expected pervasiveness of UE types in O-RAN. This change poses a higher risk to the attack surface and can be a fertile ground for new emerging threats arriving from the UE. Additionally, the IoT category which is one of the drivers of 5G and O-RAN architecture, is a device category mostly characterized by unattended use so the ability to compromise it and stay undetected for long periods is highly possible and as such extends the overall risk to the network. The source of threats on UEs are versatile and mostly rely on the software stacks that reside inside the UE. In previous cellular architectures the UE usually belonged to the category of smartphones and the option of smartphones to dynamically load applications and software is the main attack vector for compromising UEs. The path from a compromised UE onto the cellular network is diverse and involves the exploitation of vulnerabilities in the cellular services layers as well as the RF channels where a malicious UE must reside in the wireless range to carry the threat. The subtle risk of UE-based threats is the fact that the actor can be fully remote controlling the UE while the UE itself needs to be in the wireless range.

\subsubsection{Relevant Threats}
\noindent\textbf{Cellular Protocol Vulnerabilities:} Threats aiming to exploit vulnerabilities in the O-RAN and cellular protocols \cite{hussain20195greasoner,jover2019security} and their cryptographic design \cite{dehnel2018security,ferrag2018security,rupprecht2019breaking}. 

\noindent\textbf{Passive Eavesdropping:} Threats aiming to eavesdrop on the RF channel aimed at extracting sensitive information from the UE/RU communication channel such as UE location, SMS data,and others posing a major threat to confidentiality \cite{kapetanovic2015physical}.

\noindent\textbf{Radio Jamming:} Threats aiming to intentionally direct electromagnetic energy towards a radio-based communication system to disrupt or prevent signal transmission \cite{adamy2004ew}, posing a major threat to the availability of the radio channel. Previous works distinguish between two major types of jamming: threats targeting to impact the channel state information (CSI) service \cite{sodagari2015singularity,sodagari2012efficient,clancy2011efficient,miller2011vulnerabilities,vadlamani2016jamming}, and threats that impact the RF signals \cite{vadlamani2016jamming,wang2009cooperative,jorswieck2005optimal,brady2006spatially,yang2007joint,farahmand2008anti,chi2008effects,mehdi2009analysis,mukherjee2010equilibrium,asadullah2009joint}.

\noindent\textbf{Side Channel:} Threats aiming to use cellular RF protocols for achieving information leakages, such as UE tracking \cite{hong2018guti}, data leakage \cite{hussain2019privacy}, and impersonation \cite{rupprecht2020imp4gt}, posing a major threat to confidentiality and integrity.

\noindent\textbf{UE Vulnerabilities:} UE is provided by an increasing number of hardware and software vendors. Like with any software, UE software contained known and unknown vulnerabilities. Threats in this category aim to compromise the UE, and to exploit it for further impact. Those threats include mostly UE malware \cite{acar2016sok,zhou2012dissecting,becher2011mobile,kotzias2020did,li2007study,alrawi2021circle} used for gathering private information on the user, billing fraud, and more.

\noindent\textbf{Application Attacks:} UE types can vary from mobile phones to sensors, connected cars, and more. Threats in this category aim to exploit the IoT type of UE, including connected cars and sensors. Those threats include increasing UE power consumption \cite{shakhov2017energy} and harming the integrity of the IoT application \cite{hassija2019survey}.

\noindent\textbf{Side-Channel Attacks:} Threats \cite{spreitzer2017systematic} aiming to use physical properties of the UE to leak information, by using side-channel attacks.

\noindent\textbf{UE Botnets:} Threats \cite{antonakakis2017understanding,kolias2017ddos,bertino2017botnets,zhang2014iot,marzano2018evolution,edwards2016hajime} targeting a large number of UEs aiming to create a collaborative bot network (i.e. botnet) which can pose threats on the cellular network where the most popular one is the distributed denial of service (i.e. DDoS).

\noindent\textbf{DoS on Cellular Services:} Threats aiming to use similar DoS methods presented in LTE, and to adapt them to 5G. Those threats include overloading SMS control channel \cite{enck2005exploiting,racic2008exploiting,traynor2007attack}, VoLTE service and DoS on LTE terminal \cite{kim2015breaking,yu2019effects}.

\noindent\textbf{Network/transport-level DDoS flooding attacks:} Attacks utilizing TCP, UDP, ICMP and DNS protocol aimed at disrupting connectivity by exhausting bandwidth \cite{yan2015software}.

\noindent\textbf{Denial of service on the control plane:} Generating unknown packets from the data plane to the control plane can achieve denial of service when the number of packets is high \cite{shin2013attacking}. 

\noindent\textbf{Frameworks for Vulnerability Detection:} Due to the complexity of cellular networks, multiple works suggested methods to automate the process for vulnerability detection in LTE infrastructure \cite{rupprecht2016putting,hussain2018lteinspector} and protocols \cite{raza2017exposing}, \cite{kim2019touching}. Those methods, with some changes, may also be used to evaluate the O-RAN security, and to detect vulnerabilities. Threats may use previously published vulnerability detection methodologies, and to adapt them to detect vulnerabilities in O-RAN. The impact of those vulnerabilities is usually on the availability, confidentiality and integrity of the cellular network and services.

\subsection{Architectural openness}
A key technology change in O-RAN’s architecture is disaggregation and openness to a diverse supply chain both for hardware and software elements. Architectural openness has been a winning concept across the years for technologies while they have continuously depicted a unique lifecycle where at the early stages the attack surface was vastly large due to lack of control and synchronization across the entities while within the time the level of security in overall has increased in comparison to proprietary alternatives. An open architecture dictates many aspects in the technology lifecycle starting from technology acquisition, setup and deployment up to maintenance where the human factor and processes are a big part of it. From a security point of view the risk of architectural openness covers known threats from different domains that are rooted in the power openness gives the threat actors.

\noindent\textbf{Human Errors and Misconfiguration:} Misconfiguration resulting from human errors where humans can be the developers of components, integrators, engineers, and operators, represent the highest risk in open systems where configuration is the main concept for organizing the technology. Misconfigurations lead to exposed systems, incorrect access rights and exposed vulnerabilities exploited by threats \cite{nobles2018botching}. The human factor is a long-researched topic in cyber security and mitigations exist in the forms of education, awareness, verification tools and others but still it is an open issue even for long-established technologies.

\noindent\textbf{Supply Chain Software:} These threats exploit vulnerabilities in the software supply chain \cite{nist-800-161} and include exploiting or implanting vulnerabilities in software components required for the operation of the system. The vulnerability can be implanted during the authorized software’s development, packaging, or shipping stage \cite{sasaki2020security} . In recent years the category of software supply chain threats has exploded with new attacks in the wild which presented new methods for attackers to get into a dependent software package. One of the major challenges of supply chain threats, in general, is the fact they can be inactive for long durations and due to that, they can go unnoticed. 

\noindent\textbf{Supply Chain Hardware:} These threats manipulate hardware before deployment, including in the design, manufacturing, and shipping stages \cite{sasaki2020security}. These threats include the insertion of counterfeit hardware, unauthorized production, tampering, theft and insertion of malicious hardware \cite{nist-800-161}. The challenge of hardware-based supply chain threats is further exacerbated due to the fact it is near to impossible to identify malicious transplants in hardware. 

\noindent\textbf{Vulnerable Open-Source Packages:} These threats exploit vulnerabilities in open-source packages, such as Linux Bash \cite{mary2015shellshock}, SSH, and other packages. Open-source packages, particularly less mature packages, are prone to software bugs, however patching open-source packages after deployment is rarely done. Attackers are enjoying the open source disclosure cycle where new vulnerabilities are reported and there is a gap in between the knowledge on the vulnerability is public to the moment is system is patched and within that time window the majority of the attacks take place.

\noindent\textbf{API Exploitation:} These threats exploit application programming interface (API) interfaces. API interfaces expose software functionality to authorized interactions with other components to enable interoperability and composability. However, if the interface is not secure via proper authorization, authentication and input sanitation, those APIs can be used to manipulate the software \cite{zhang2018code}, cause denial of service, and even execute unauthorized code. 

\subsection{Cloud and Virtualization}
The O-RAN architecture is highly dependent on cloud infrastructure as part of its openness concept. Cloud infrastructure streamlines application development, deployment and monitoring and allows multiple functionalities on a single logical unit. The stack of cloud on O-RAN consists on lower layer virtualization and application execution environment in the form of containers where Kubernetes is the go to container orchestration platform. The IT world has shifted into cloud-based computing in the recent decade where clouds appear both in the on-premise data center as well as in public clouds. Due to the extensive growth in the cloud category, there was also an explosion in the number of threats and attacks targeting different aspects of the cloud stack and these risks are highly transferable to the O-RAN landscape.

\noindent\textbf{Co-Hosted Application Side Channels:} These threats exploit scenarios in which multiple applications are executed on the same hardware where the attacker is exploiting CPU-level vulnerabilities to extract sensitive data from memory. As O-RAN is expected to run commodity cloud and hardware infrastructure this threat is highly relevant. The following research has demonstrated the possibility of data leakage between applications \cite{kocher2019spectre}. 

\noindent\textbf{Image Manipulation:} In virtualized environments, applications are stored and executed from a binary image file, regardless of whether it is a virtual machine (VM) or a docker file. Some threats target target those image files, which try to exploit vulnerabilities during image creation or execution \cite{almorsy2016analysis,pattaranantakul2019moving,shu2017study,zhang2012deep,ristenpart2009hey,lin2018measurement}. In addition, some threats attempt to ’break’ the application’s execution, to influence applications that are outside the boundary of the threat’s malicious image. This threat is considered under the general category of supply chain threats in the case of manipulating the image at rest as it manipulates dependent assets before execution on the target environment in a fashion that is difficult to detect once it reaches the runtime environment. 

\noindent\textbf{Guest-to-Guest Attacks:} In virtualized environments, applications share the same resources and may share storage, a CPU, or a host OS. Several threats target those shared resources, in an attempt to communicate with other guest applications \cite{ristenpart2009hey}, inject malicious execution code between virtual machines \cite{sgandurra2016evolution}, leak information through side-channel attacks \cite{zhang2012cross,irazoqui2015s,zhang2014cross,liu2015last}, and perform DoS attacks on other guest applications \cite{Liyanage2017}. These type of threats when considering their relevancy to O-RAN which is not a public cloud where there is little knowledge on other running entity are targeted attacks as the attacker need intimate knowledge on the other tenants running on the same infrastructure. Furthermore, the type of vulnerabilities, in these domains are hard to identify as these are highly popular platforms that are scrutinized continuously in other domains. 

\noindent\textbf{Guest-to-Hypervisor Attacks:} These threats are aimed at harming the integrity of the host OS or the hypervisor \cite{zhang2012deep,huang2012security} and include seamlessly moving a virtual machine from one hardware component to another \cite{atya2017stalling} and performing a DoS attack on the host or other guest applications \cite{Liyanage2017}. The type of threats and vulnerabilities in this category are similar to the ones in the Guest-to-Guest category in terms of the complexity of uncovering.

\noindent\textbf{Inconsistent Security Policies:} These threats exploit the misconfiguration of the complex security policies of cloud-based assets. This includes exploiting the misconfiguration of cloud databases \cite{ferrari2020nosql}, storage \cite{continella2018there}, and computation units \cite{Liyanage2017}. This threat category is under the general misconfiguration category while in cloud it got special attention as the assets that can be targeted are naturally in a perimeter-less environment and such attackers have many opportunities to exploit them. Within O-RAN which is a controlled cloud environment the risk of these threats is lower as there is a physical perimeter surrounding the assets.

\noindent\textbf{Exploiting Public-Facing Applications:} These threats exploit public-facing services \cite{stuttard2011web} provided by the cloud. This includes APIs exposed via Web services, database manipulation through a legitimate interface (e.g., SQL injections \cite{singh2016sql}), as well as other services with exposed interfaces. 

\noindent\textbf{Trusted Connected Third Parties:} Cloud systems depend on trusted relationships with third parties, including IT service contractors, managed security providers, and infrastructure contractors \cite{stuttard2011web}. These threats are aimed at exploiting those relationships to gain access to protected cloud assets, by exploiting the management protocol and valid account holders. 

\noindent\textbf{Improper SSL/TLS Configuration:} The secure Socket Layer (SSL) and Transport Layer Security (TLS) protocols provide end-to-end secure communication over the Internet. The details of the SSL/TLS protocol are complex with many security configurations and details. To ease their use by developers, these details are encapsulated in open-source SSL/TLS libraries. However, incorrect use of those libraries can lead to security problems, such as man-in-the-middle attacks \cite{he2015vetting} and security degradation, as described by \cite{clark2013sok}. 

\noindent\textbf{Rogue Fog Nodes:} When a cloud node is exploited and taken by an attacker command and control, that node can become an active rogue element which can be used for intercepting traffic or launching higher privilege attacks on other nodes or other components \cite{olimid20205g}. This scenario depends on the ability of an actor to actively control the malicious node and within O-RAN controlled cloud and taking into account proper access measures the transferability of such attack to O-RAN is low.

\subsection{Machine Learning}
One of the core ideas in O-RAN is to enable autonomous network resource management utilizing machine learning algorithms where the algorithms are confined to human-	made policies. The resource optimization ML inference capabilities are designated to run inside the near-RT RIC where the data pipeline can be executed in the SMO or any other environment. Machine Learning algorithms are highly sensitive to a whole genre of adversarial ML threats where main pathways for actor are via the data used for training a model, the model itself, the data used for inference and the software packaging surrounding these elements. Adversarial ML threats are an emerging topic and there is a growing body of knowledge articulating different threats and attack techniques while in general the transferability across domain seems very high. The main risk in the world of adversarial ML is the immaturity of countermeasures including the ability to monitor, detect and prevent such threats. Relying on ML for the complete network management is a must from an architectural point of view as the complexity of O-RAN networks in different deployment scenarios can be immense and beyond a human capability to understand the full picture and the interaction in between decisions. Still, it is heavy duty and as such it requires a deep understanding of the threats and planning for proper mitigations during the model training stages as well as inference. 
Another aspect of ML inside O-RAN is the use of third-party or operator applications and services that rely on ML capabilities and the list of threat genres articulated here are highly relevant to these as well though it is difficult to analyze the level of threat in this case as there are no specifications for such capabilities. 
Therefore, threats that target general machine learning systems have become a major threat to O-RAN. 
Threats to machine learning systems are surveyed in \cite{papernot2018sok} which we treat as a gateway for the world of adversarial threats. The level of detail in this section is low as the topic is highly complex and is detailed in a standalone report of ML Adversarial Threats in O-RAN. Furthermore, we focus only on the system-level ML capabilities and not general ML capabilities provided by third parties.

\noindent\textbf{Misprediction:} These threats \cite{he2019towards,wang2018stealing,suomalainen2020machine} are aimed at harming the integrity of the algorithm, by manipulating its decision-making to misclassify or mispredict based on specific incoming data. These types of attacks have intimate knowledge on blind spots in the way the model is structured and are manipulating the incoming data used for inference to achieve a different result than originally intended. In O-RAN, those threats can have a significant impact on all functionalities related to network resource management including traffic shaping, load balancing and general optimization algorithms. 

\noindent\textbf{Membership Attacks:} These threats \cite{he2019towards,wang2018stealing,suomalainen2020machine} aim to determine whether a specific data point was part of the model’s training dataset. The attack risk lies in privacy where an attacker can expose the fact a certain piece of information exists within a certain database (i.e. certain VIP characteristics).  

\noindent\textbf{Training Data Extraction:} These threats \cite{he2019towards,wang2018stealing,suomalainen2020machine} are aimed at extracting the data used for training and building the model. Unlike the membership attack which is aimed at identifying the existence of a specific data point in a training set, in a training data extraction attack the attacker can recover a complete data point from scratch. This risk is on the privacy aspect of the system while not necessarily limited to that as the extracted data can be used in later-on lateral attacks.

\noindent\textbf{Model Extraction:} These threats \cite{he2019towards,wang2018stealing,suomalainen2020machine} are aimed at extracting the machine learning model via the interface provided to the model, without prior knowledge. Extracted models can pose risks ranging from IP theft up to privacy breaching in the case of recovering training set elements and most importantly it can be used for developing accurate misprediction attacks.

\subsection{5G Architecture}
O-RAN as a 5G compatible RAN architecture is aimed to support multiple QoS levels for services via slices using the same infrastructure \cite{nencioni2018orchestration}. Along with its benefits, network slicing faces several new threats that are general to 5G architectures and applies to O-RAN as well. Those threats are surveyed in \cite{olimid20205g}.

\noindent\textbf{Slice Lifecycle Security:} O-RAN slices have four phases of operation: preparation, instantiation, runtime, and decommissioning. Each of those phases faces different types of threats, which have been described in detail in \cite{olimid20205g}. Those threats include changes to the network slice template, configuration changes, and information leakage. 

\noindent\textbf{Intra-Slice Security:} These threats target a specific slice and do not have any influence on other slices. These threats include slice service interface exploitation, DoS attacks from UE, and API exploitation of the slice manager. 

\noindent\textbf{Inter-Slice Security:} These threats target a specific slice and influence other slices. These threats include unauthorized communication from a low-security slice to a more secure one and DoS on the slice’s manager service. 

\section{\label{sec:discussion} Discussion}
This section describes the main risks to O-RAN identified during the analysis which require special attention: 

\subsection{Increased Attack Surface due to Ecosystem.}
In previous RAN proprietary architectures, the number of parties that were involved in supplying, developing, maintaining, and operating the technology was in magnitude smaller than the ones contemplated in O-RAN. Furthermore, the number of participants who have direct access to the assets eventually running within the RAN is greater as well than previous generations. This state of the ecosystem creates a vast attack surface that is not fully controllable. Such cases of complex ecosystems in other domains such as large web services are managed via mature processes, policies, and guidelines and there is a need to adopt that modus operandi as part of launching O-RAN ventures. A major topic that will be highly prominent in such a world will be the supply chain risk which will require special attention in terms of tools, processes, and guidelines.

\subsection{Increased Attack Surface due to Diversity of UEs.}
A large number of threats on previous cellular architectures targeted the UE. The impact of those threats was on the users (mainly on confidentiality) and the core network (availability).
In O-RAN, there is a dramatic increase in the types of UE, and the UE is no longer limited to cellular phones and includes connected cars, smart sensors, smart meters, disposable sensors and more. In general, the additional types of UE are less protected than mobile phones, and we expect to see them targeted and potentially exploited towards targeting the cellular network. The concern for the security of the new UE categories is not limited only to the network operator, it is a concern for the whole industry surrounding these new devices and as such the level of security is expected to improve regardless of operators’ efforts. 

\subsection{Increased Attack Surface due to Diversity of 3rd Party Applications.}
One or maybe the major opportunity within 5G RAN networks is the ability to offload compute from UEs into applications running inside the O-RAN cloud environment based on latency requirements. Applications such as face recognition, passenger recognition, smart factory optimization and others eventually will have a component that will run inside the RAN and will serve their customers. This modus operandi introduces a plethora of challenges similar to the ones experienced by operators of public clouds and requires the adoption of guidelines, policies and tools to secure the platform on one hand and to ease the development of more value-added services on the other.

\subsection{Increased Attack Surface due to Integrated Machine Learning.}
The integration of ML into the resource management part of the RAN will be a one-way path where the automation will be required to be overarching and complete to operate properly. From a technology point of view that means dependency on millions of decisions taken every second by different ML models and a larger and complex process of models training, verification, deployment, monitoring and refinement. Machine learning models as a technological paradigm exhibit obscure functionality where it is very difficult to understand how the models take decisions and what is the impact of those decisions and that obscurity overflows into the early stages of turning raw data into mathematical models. This level of obscurity is an open attack surface area especially considering the inability to control the data sources that are used for making those decisions. Another large challenge for the introduction of machine learning is the lack of mature countermeasures industry and principles and that would require specific research and development work to fortify the models that are going to be deployed.

\section{Conclusion}
The attack surface of O-RAN based on its architecture is vastly larger than the previous proprietary RAN architectures due to the openness of the architecture as well due to the higher requirements from the platform in terms of the number of involved parties which is driven mostly by 5G requirements and less due to specific decisions related to O-RAN.
It may be more accurate to say that the attack surface of the RAN is now clearer vs. the proprietary implementations which hide behind obscurities. The fact the attack surface is clearer contributes dramatically to the ability to defend it and since many of the concepts in O-RAN are borrowed from mature technological paradigms such as cloud there is a high transferability potential for countermeasures to be useful within O-RAN. The openness of O-RAN in general offers a great promise for secure radio access networks in the future as it has been proven that open systems which enjoy massive scrutiny from the community, at the first stages in the lifecycle may be more vulnerable than proprietary ones but quite fast they become robust in a parallel level to the proprietary ones and keeps on improving without being confined to the efforts and motivation of a specific vendor. From a capacity point of view the massive number of requirements from 5G networks it seems that going in the open direction which relies on contribution from many parties is the only path ahead to be able to unleash innovation on hand and keep systems secure. This paper which is possible thanks to the open specification is a great example of the value of openness in terms of contribution to the security level of the RAN.

\bibliographystyle{IEEEtranS}
\bibliography{main.bib}

\begin{thebibliography}{100}
\providecommand{\url}[1]{#1}
\csname url@samestyle\endcsname
\providecommand{\newblock}{\relax}
\providecommand{\bibinfo}[2]{#2}
\providecommand{\BIBentrySTDinterwordspacing}{\spaceskip=0pt\relax}
\providecommand{\BIBentryALTinterwordstretchfactor}{4}
\providecommand{\BIBentryALTinterwordspacing}{\spaceskip=\fontdimen2\font plus
\BIBentryALTinterwordstretchfactor\fontdimen3\font minus
  \fontdimen4\font\relax}
\providecommand{\BIBforeignlanguage}[2]{{%
\expandafter\ifx\csname l@#1\endcsname\relax
\typeout{** WARNING: IEEEtranS.bst: No hyphenation pattern has been}%
\typeout{** loaded for the language `#1'. Using the pattern for}%
\typeout{** the default language instead.}%
\else
\language=\csname l@#1\endcsname
\fi
#2}}
\providecommand{\BIBdecl}{\relax}
\BIBdecl

\bibitem{ciscowebsite}
``Cisco 2018 annual report,''
  \url{https://www.cisco.com/c/dam/en\_us/about/annual-report/2018-annual-report-full.pdf},
  accessed: 2021-08-04.

\bibitem{acar2016sok}
Y.~Acar, M.~Backes, S.~Bugiel, S.~Fahl, P.~McDaniel, and M.~Smith, ``Sok:
  Lessons learned from android security research for appified software
  platforms,'' in \emph{2016 IEEE Symposium on Security and Privacy
  (SP)}.\hskip 1em plus 0.5em minus 0.4em\relax IEEE, 2016, pp. 433--451.

\bibitem{adamy2004ew}
D.~Adamy, \emph{EW 102: a second course in electronic warfare}.\hskip 1em plus
  0.5em minus 0.4em\relax Artech House, 2004.

\bibitem{alliance2018ran}
O.~R. Alliance, ``O-ran: towards an open and smart ran,'' \emph{white paper,
  October}, 2018.

\bibitem{alliance2020ran}
O.~Alliance, ``O-ran use cases and deployment scenarios,'' \emph{White Paper},
  2020.

\bibitem{alliance2021ran}
------, ``O-ran architecture description,'' \emph{O-RAN. WG1.
  O-RAN-Architecture-Description-v03. 00, Technical Specification}, 2021.

\bibitem{almorsy2016analysis}
M.~Almorsy, J.~Grundy, and I.~M{\"u}ller, ``An analysis of the cloud computing
  security problem,'' \emph{arXiv preprint arXiv:1609.01107}, 2016.

\bibitem{alnoman2017towards}
A.~Alnoman and A.~Anpalagan, ``Towards the fulfillment of 5g network
  requirements: technologies and challenges,'' \emph{Telecommunication
  Systems}, vol.~65, no.~1, pp. 101--116, 2017.

\bibitem{alrawi2021circle}
O.~Alrawi, C.~Lever, K.~Valakuzhy, K.~Snow, F.~Monrose, M.~Antonakakis
  \emph{et~al.}, ``The circle of life: A large-scale study of the iot malware
  lifecycle,'' in \emph{30th $\{$USENIX$\}$ Security Symposium ($\{$USENIX$\}$
  Security 21)}, 2021.

\bibitem{antonakakis2017understanding}
M.~Antonakakis, T.~April, M.~Bailey, M.~Bernhard, E.~Bursztein, J.~Cochran,
  Z.~Durumeric, J.~A. Halderman, L.~Invernizzi, M.~Kallitsis \emph{et~al.},
  ``Understanding the mirai botnet,'' in \emph{26th $\{$USENIX$\}$ security
  symposium ($\{$USENIX$\}$ Security 17)}, 2017, pp. 1093--1110.

\bibitem{asadullah2009joint}
M.~G. Asadullah and G.~L. Stuber, ``Joint iterative channel estimation and
  soft-chip combining for a mimo mc-cdma anti-jam system,'' \emph{IEEE
  transactions on communications}, vol.~57, no.~4, pp. 1068--1078, 2009.

\bibitem{ateya2018study}
A.~A. Ateya, A.~Muthanna, M.~Makolkina, and A.~Koucheryavy, ``Study of 5g
  services standardization: specifications and requirements,'' in \emph{2018
  10th International Congress on Ultra Modern Telecommunications and Control
  Systems and Workshops (ICUMT)}.\hskip 1em plus 0.5em minus 0.4em\relax IEEE,
  2018, pp. 1--6.

\bibitem{atya2017stalling}
A.~Atya, A.~Aqil, K.~Khalil, Z.~Qian, S.~V. Krishnamurthy, and T.~F. La~Porta,
  ``Stalling live migrations on the cloud,'' in \emph{11th $\{$USENIX$\}$
  Workshop on Offensive Technologies ($\{$WOOT$\}$ 17)}, 2017.

\bibitem{balasubramanian2021ric}
B.~Balasubramanian, E.~S. Daniels, M.~Hiltunen, R.~Jana, K.~Joshi, R.~Sivaraj,
  T.~X. Tran, and C.~Wang, ``Ric: A ran intelligent controller platform for
  ai-enabled cellular networks,'' \emph{IEEE Internet Computing}, vol.~25,
  no.~2, pp. 7--17, 2021.

\bibitem{becher2011mobile}
M.~Becher, F.~C. Freiling, J.~Hoffmann, T.~Holz, S.~Uellenbeck, and C.~Wolf,
  ``Mobile security catching up? revealing the nuts and bolts of the security
  of mobile devices,'' in \emph{2011 IEEE Symposium on Security and
  Privacy}.\hskip 1em plus 0.5em minus 0.4em\relax IEEE, 2011, pp. 96--111.

\bibitem{bertino2017botnets}
E.~Bertino and N.~Islam, ``Botnets and internet of things security,''
  \emph{Computer}, vol.~50, no.~2, pp. 76--79, 2017.

\bibitem{bonati2020open}
L.~Bonati, M.~Polese, S.~D’Oro, S.~Basagni, and T.~Melodia, ``Open,
  programmable, and virtualized 5g networks: State-of-the-art and the road
  ahead,'' \emph{Computer Networks}, vol. 182, p. 107516, 2020.

\bibitem{nist-800-161}
J.~Boyens, C.~Paulsen, R.~Moorthy, and N.~Bartol,
  ``\BIBforeignlanguage{en}{Supply chain risk management practices for federal
  information systems and organizations},'' 2015-04-08 2015.

\bibitem{brady2006spatially}
M.~H. Brady, M.~Mohseni, and J.~M. Cioffi, ``Spatially-correlated jamming in
  gaussian multiple access and broadcast channels,'' in \emph{2006 40th Annual
  Conference on Information Sciences and Systems}.\hskip 1em plus 0.5em minus
  0.4em\relax IEEE, 2006, pp. 1635--1639.

\bibitem{chaudhary2019c}
J.~K. Chaudhary, A.~Kumar, J.~Bartelt, and G.~Fettweis, ``C-ran employing xran
  functional split: Complexity analysis for 5g nr remote radio unit,'' in
  \emph{2019 European Conference on Networks and Communications (EuCNC)}.\hskip
  1em plus 0.5em minus 0.4em\relax IEEE, 2019, pp. 580--585.

\bibitem{chi2008effects}
D.~W. Chi and P.~Das, ``Effects of jammer and nonlinear amplifiers in mimo-ofdm
  with application to 802.11 n wlan,'' in \emph{MILCOM 2008-2008 IEEE Military
  Communications Conference}.\hskip 1em plus 0.5em minus 0.4em\relax IEEE,
  2008, pp. 1--8.

\bibitem{clancy2011efficient}
T.~C. Clancy, ``Efficient ofdm denial: Pilot jamming and pilot nulling,'' in
  \emph{2011 IEEE International Conference on Communications (ICC)}.\hskip 1em
  plus 0.5em minus 0.4em\relax IEEE, 2011, pp. 1--5.

\bibitem{clark2013sok}
J.~Clark and P.~C. Van~Oorschot, ``Sok: Ssl and https: Revisiting past
  challenges and evaluating certificate trust model enhancements,'' in
  \emph{2013 IEEE Symposium on Security and Privacy}.\hskip 1em plus 0.5em
  minus 0.4em\relax IEEE, 2013, pp. 511--525.

\bibitem{continella2018there}
A.~Continella, M.~Polino, M.~Pogliani, and S.~Zanero, ``There's a hole in that
  bucket! a large-scale analysis of misconfigured s3 buckets,'' in
  \emph{Proceedings of the 34th Annual Computer Security Applications
  Conference}, 2018, pp. 702--711.

\bibitem{dehnel2018security}
M.~Dehnel-Wild and C.~Cremers, ``Security vulnerability in 5g-aka draft,''
  \emph{Department of Computer Science, University of Oxford, Tech. Rep}, pp.
  14--37, 2018.

\bibitem{edwards2016hajime}
S.~Edwards and I.~Profetis, ``Hajime: Analysis of a decentralized internet worm
  for iot devices,'' \emph{Rapidity Networks}, vol.~16, pp. 1--18, 2016.

\bibitem{elayoubi20195g}
S.~E. Elayoubi, S.~B. Jemaa, Z.~Altman, and A.~Galindo-Serrano, ``5g ran
  slicing for verticals: Enablers and challenges,'' \emph{IEEE Communications
  Magazine}, vol.~57, no.~1, pp. 28--34, 2019.

\bibitem{enck2005exploiting}
W.~Enck, P.~Traynor, P.~McDaniel, and T.~La~Porta, ``Exploiting open
  functionality in sms-capable cellular networks,'' in \emph{Proceedings of the
  12th ACM conference on Computer and communications security}, 2005, pp.
  393--404.

\bibitem{farahmand2008anti}
S.~Farahmand, A.~Cano, and G.~B. Giannakis, ``Anti-jam distributed mimo
  decoding using wireless sensor networks,'' in \emph{2008 IEEE International
  Conference on Acoustics, Speech and Signal Processing}.\hskip 1em plus 0.5em
  minus 0.4em\relax IEEE, 2008, pp. 2257--2260.

\bibitem{ferrag2018security}
M.~A. Ferrag, L.~Maglaras, A.~Argyriou, D.~Kosmanos, and H.~Janicke, ``Security
  for 4g and 5g cellular networks: A survey of existing authentication and
  privacy-preserving schemes,'' \emph{Journal of Network and Computer
  Applications}, vol. 101, pp. 55--82, 2018.

\bibitem{ferrari2020nosql}
D.~Ferrari, M.~Carminati, M.~Polino, and S.~Zanero, ``Nosql breakdown: A
  large-scale analysis of misconfigured nosql services,'' in \emph{Annual
  Computer Security Applications Conference}, 2020, pp. 567--581.

\bibitem{global2017state}
G.~S. for Mobile~communications Association \emph{et~al.}, ``State of the
  industry report on mobile money,'' 2017.

\bibitem{habibi2018structure}
M.~A. Habibi, B.~Han, M.~Nasimi, and H.~D. Schotten, ``The structure of service
  level agreement of slice-based 5g network,'' \emph{arXiv preprint
  arXiv:1806.10426}, 2018.

\bibitem{hassija2019survey}
V.~Hassija, V.~Chamola, V.~Saxena, D.~Jain, P.~Goyal, and B.~Sikdar, ``A survey
  on iot security: application areas, security threats, and solution
  architectures,'' \emph{IEEE Access}, vol.~7, pp. 82\,721--82\,743, 2019.

\bibitem{he2015vetting}
B.~He, V.~Rastogi, Y.~Cao, Y.~Chen, V.~Venkatakrishnan, R.~Yang, and Z.~Zhang,
  ``Vetting ssl usage in applications with sslint,'' in \emph{2015 IEEE
  Symposium on Security and Privacy}.\hskip 1em plus 0.5em minus 0.4em\relax
  IEEE, 2015, pp. 519--534.

\bibitem{he2019towards}
Y.~He, G.~Meng, K.~Chen, X.~Hu, and J.~He, ``Towards security threats of deep
  learning systems: A survey,'' \emph{arXiv preprint arXiv:1911.12562}, 2019.

\bibitem{hong2018guti}
B.~Hong, S.~Bae, and Y.~Kim, ``Guti reallocation demystified: Cellular location
  tracking with changing temporary identifier.'' in \emph{NDSS}, 2018.

\bibitem{huang2012security}
Y.-L. Huang, B.~Chen, M.-W. Shih, and C.-Y. Lai, ``Security impacts of
  virtualization on a network testbed,'' in \emph{2012 IEEE Sixth International
  Conference on Software Security and Reliability}.\hskip 1em plus 0.5em minus
  0.4em\relax IEEE, 2012, pp. 71--77.

\bibitem{hussain2018lteinspector}
S.~Hussain, O.~Chowdhury, S.~Mehnaz, and E.~Bertino, ``Lteinspector: A
  systematic approach for adversarial testing of 4g lte,'' in \emph{Network and
  Distributed Systems Security (NDSS) Symposium 2018}, 2018.

\bibitem{hussain2019privacy}
S.~R. Hussain, M.~Echeverria, O.~Chowdhury, N.~Li, and E.~Bertino, ``Privacy
  attacks to the 4g and 5g cellular paging protocols using side channel
  information,'' \emph{Network and Distributed Systems Security (NDSS)
  Symposium2019}, 2019.

\bibitem{hussain20195greasoner}
S.~R. Hussain, M.~Echeverria, I.~Karim, O.~Chowdhury, and E.~Bertino,
  ``5greasoner: A property-directed security and privacy analysis framework for
  5g cellular network protocol,'' in \emph{Proceedings of the 2019 ACM SIGSAC
  Conference on Computer and Communications Security}, 2019, pp. 669--684.

\bibitem{irazoqui2015s}
G.~Irazoqui, T.~Eisenbarth, and B.~Sunar, ``Sa: A shared cache attack that
  works across cores and defies vm sandboxing--and its application to aes,'' in
  \emph{2015 IEEE Symposium on Security and Privacy}.\hskip 1em plus 0.5em
  minus 0.4em\relax IEEE, 2015, pp. 591--604.

\bibitem{jiang2017overview}
D.~Jiang and G.~Liu, ``An overview of 5g requirements,'' \emph{5G Mobile
  Communications}, pp. 3--26, 2017.

\bibitem{jorswieck2005optimal}
E.~Jorswieck, H.~Boche, and M.~Weckerle, ``Optimal transmitter and jamming
  strategies in gaussian mimo channels,'' in \emph{2005 IEEE 61st Vehicular
  Technology Conference}, vol.~2.\hskip 1em plus 0.5em minus 0.4em\relax IEEE,
  2005, pp. 978--982.

\bibitem{jover2013security}
R.~P. Jover, ``Security attacks against the availability of lte mobility
  networks: Overview and research directions,'' in \emph{2013 16th
  international symposium on wireless personal multimedia communications
  (WPMC)}.\hskip 1em plus 0.5em minus 0.4em\relax IEEE, 2013, pp. 1--9.

\bibitem{jover2019security}
R.~P. Jover and V.~Marojevic, ``Security and protocol exploit analysis of the
  5g specifications,'' \emph{IEEE Access}, vol.~7, pp. 24\,956--24\,963, 2019.

\bibitem{kapetanovic2015physical}
D.~Kapetanovic, G.~Zheng, and F.~Rusek, ``Physical layer security for massive
  mimo: An overview on passive eavesdropping and active attacks,'' \emph{IEEE
  Communications Magazine}, vol.~53, no.~6, pp. 21--27, 2015.

\bibitem{kim2015breaking}
H.~Kim, D.~Kim, M.~Kwon, H.~Han, Y.~Jang, D.~Han, T.~Kim, and Y.~Kim,
  ``Breaking and fixing volte: Exploiting hidden data channels and
  mis-implementations,'' in \emph{Proceedings of the 22nd ACM SIGSAC Conference
  on Computer and Communications Security}, 2015, pp. 328--339.

\bibitem{kim2019touching}
H.~Kim, J.~Lee, E.~Lee, and Y.~Kim, ``Touching the untouchables: Dynamic
  security analysis of the lte control plane,'' in \emph{2019 IEEE Symposium on
  Security and Privacy (SP)}.\hskip 1em plus 0.5em minus 0.4em\relax IEEE,
  2019, pp. 1153--1168.

\bibitem{kocher2019spectre}
P.~Kocher, J.~Horn, A.~Fogh, D.~Genkin, D.~Gruss, W.~Haas, M.~Hamburg, M.~Lipp,
  S.~Mangard, T.~Prescher \emph{et~al.}, ``Spectre attacks: Exploiting
  speculative execution,'' in \emph{2019 IEEE Symposium on Security and Privacy
  (SP)}.\hskip 1em plus 0.5em minus 0.4em\relax IEEE, 2019, pp. 1--19.

\bibitem{kolias2017ddos}
C.~Kolias, G.~Kambourakis, A.~Stavrou, and J.~Voas, ``Ddos in the iot: Mirai
  and other botnets,'' \emph{Computer}, vol.~50, no.~7, pp. 80--84, 2017.

\bibitem{kotzias2020did}
P.~Kotzias, J.~Caballero, and L.~Bilge, ``How did that get in my phone?
  unwanted app distribution on android devices,'' \emph{arXiv preprint
  arXiv:2010.10088}, 2020.

\bibitem{li2007study}
W.-J. Li, S.~Stolfo, A.~Stavrou, E.~Androulaki, and A.~D. Keromytis, ``A study
  of malcode-bearing documents,'' in \emph{International conference on
  detection of intrusions and malware, and vulnerability assessment}.\hskip 1em
  plus 0.5em minus 0.4em\relax Springer, 2007, pp. 231--250.

\bibitem{lin2018measurement}
X.~Lin, L.~Lei, Y.~Wang, J.~Jing, K.~Sun, and Q.~Zhou, ``A measurement study on
  linux container security: Attacks and countermeasures,'' in \emph{Proceedings
  of the 34th Annual Computer Security Applications Conference}, 2018, pp.
  418--429.

\bibitem{liu2015last}
F.~Liu, Y.~Yarom, Q.~Ge, G.~Heiser, and R.~B. Lee, ``Last-level cache
  side-channel attacks are practical,'' in \emph{2015 IEEE symposium on
  security and privacy}.\hskip 1em plus 0.5em minus 0.4em\relax IEEE, 2015, pp.
  605--622.

\bibitem{liu20165g}
G.~Liu and D.~Jiang, ``5g: Vision and requirements for mobile communication
  system towards year 2020,'' \emph{Chinese Journal of Engineering}, vol. 2016,
  no. 2016, p.~8, 2016.

\bibitem{Liyanage2017}
M.~{Liyanage}, I.~{Ahmad}, A.~B. {Abro}, A.~{Gurtov}, and M.~{Ylianttila},
  \emph{Cloud and MEC Security}, 2017, pp. 373--397.

\bibitem{mary2015shellshock}
C.~Mary, ``Shellshock attack on linux systems--bash,'' \emph{International
  Research Journal of Engineering and Technology}, vol.~2, no.~8, pp.
  1322--1325, 2015.

\bibitem{marzano2018evolution}
A.~Marzano, D.~Alexander, O.~Fonseca, E.~Fazzion, C.~Hoepers,
  K.~Steding-Jessen, M.~H. Chaves, {\'I}.~Cunha, D.~Guedes, and W.~Meira, ``The
  evolution of bashlite and mirai iot botnets,'' in \emph{2018 IEEE Symposium
  on Computers and Communications (ISCC)}.\hskip 1em plus 0.5em minus
  0.4em\relax IEEE, 2018, pp. 00\,813--00\,818.

\bibitem{mattisson2018overview}
S.~Mattisson, ``An overview of 5g requirements and future wireless networks:
  Accommodating scaling technology,'' \emph{IEEE Solid-State Circuits
  Magazine}, vol.~10, no.~3, pp. 54--60, 2018.

\bibitem{mavoungou2016survey}
S.~Mavoungou, G.~Kaddoum, M.~Taha, and G.~Matar, ``Survey on threats and
  attacks on mobile networks,'' \emph{IEEE Access}, vol.~4, pp. 4543--4572,
  2016.

\bibitem{mehdi2009analysis}
H.~Mehdi, K.~C. Teh, and K.~H. Li, ``Analysis of mimo band-limited ds-cdma
  systems in the presence of multitone jamming over generalized-$ k $ fading
  channels,'' \emph{IEEE transactions on vehicular technology}, vol.~58, no.~7,
  pp. 3825--3829, 2009.

\bibitem{miller2011vulnerabilities}
R.~Miller and W.~Trappe, ``On the vulnerabilities of csi in mimo wireless
  communication systems,'' \emph{IEEE Transactions on mobile Computing},
  vol.~11, no.~8, pp. 1386--1398, 2011.

\bibitem{mukherjee2010equilibrium}
A.~Mukherjee and A.~L. Swindlehurst, ``Equilibrium outcomes of dynamic games in
  mimo channels with active eavesdroppers,'' in \emph{2010 IEEE International
  Conference on Communications}.\hskip 1em plus 0.5em minus 0.4em\relax IEEE,
  2010, pp. 1--5.

\bibitem{nencioni2018orchestration}
G.~Nencioni, R.~G. Garroppo, A.~J. Gonzalez, B.~E. Helvik, and G.~Procissi,
  ``Orchestration and control in software-defined 5g networks: research
  challenges,'' \emph{Wireless communications and mobile computing}, vol. 2018,
  2018.

\bibitem{nobles2018botching}
C.~Nobles \emph{et~al.}, ``Botching human factors in cybersecurity in business
  organizations,'' \emph{HOLISTICA--Journal of Business and Public
  Administration}, vol.~9, no.~3, pp. 71--88, 2018.

\bibitem{norp20185g}
T.~Norp, ``5g requirements and key performance indicators,'' \emph{Journal of
  ICT Standardization}, pp. 15--30, 2018.

\bibitem{olimid20205g}
R.~F. Olimid and G.~Nencioni, ``5g network slicing: a security overview,''
  \emph{IEEE Access}, vol.~8, pp. 99\,999--100\,009, 2020.

\bibitem{papernot2018sok}
N.~Papernot, P.~McDaniel, A.~Sinha, and M.~P. Wellman, ``Sok: Security and
  privacy in machine learning,'' in \emph{2018 IEEE European Symposium on
  Security and Privacy (EuroS\&P)}.\hskip 1em plus 0.5em minus 0.4em\relax
  IEEE, 2018, pp. 399--414.

\bibitem{parkvall2017nr}
S.~Parkvall, E.~Dahlman, A.~Furuskar, and M.~Frenne, ``Nr: The new 5g radio
  access technology,'' \emph{IEEE Communications Standards Magazine}, vol.~1,
  no.~4, pp. 24--30, 2017.

\bibitem{parvez2018survey}
I.~Parvez, A.~Rahmati, I.~Guvenc, A.~I. Sarwat, and H.~Dai, ``A survey on low
  latency towards 5g: Ran, core network and caching solutions,'' \emph{IEEE
  Communications Surveys \& Tutorials}, vol.~20, no.~4, pp. 3098--3130, 2018.

\bibitem{pattaranantakul2019moving}
M.~Pattaranantakul, ``Moving towards software-defined security in the era of
  nfv and sdn,'' Ph.D. dissertation, Universit{\'e} Paris-Saclay, 2019.

\bibitem{patwary2020authentication}
A.~A.-N. Patwary, A.~Fu, R.~K. Naha, S.~K. Battula, S.~Garg, M.~A.~K. Patwary,
  and E.~Aghasian, ``Authentication, access control, privacy, threats and trust
  management towards securing fog computing environments: A review,''
  \emph{arXiv preprint arXiv:2003.00395}, 2020.

\bibitem{racic2008exploiting}
R.~Racic, D.~Ma, H.~Chen, and X.~Liu, ``Exploiting opportunistic scheduling in
  cellular data networks.'' in \emph{NDSS}.\hskip 1em plus 0.5em minus
  0.4em\relax Citeseer, 2008.

\bibitem{raza2017exposing}
M.~T. Raza, F.~M. Anwar, and S.~Lu, ``Exposing lte security weaknesses at
  protocol inter-layer, and inter-radio interactions,'' in \emph{International
  Conference on Security and Privacy in Communication Systems}.\hskip 1em plus
  0.5em minus 0.4em\relax Springer, 2017, pp. 312--338.

\bibitem{ristenpart2009hey}
T.~Ristenpart, E.~Tromer, H.~Shacham, and S.~Savage, ``Hey, you, get off of my
  cloud: exploring information leakage in third-party compute clouds,'' in
  \emph{Proceedings of the 16th ACM conference on Computer and communications
  security}, 2009, pp. 199--212.

\bibitem{rost2017network}
P.~Rost, C.~Mannweiler, D.~S. Michalopoulos, C.~Sartori, V.~Sciancalepore,
  N.~Sastry, O.~Holland, S.~Tayade, B.~Han, D.~Bega \emph{et~al.}, ``Network
  slicing to enable scalability and flexibility in 5g mobile networks,''
  \emph{IEEE Communications magazine}, vol.~55, no.~5, pp. 72--79, 2017.

\bibitem{rupprecht2016putting}
D.~Rupprecht, K.~Jansen, and C.~P{\"o}pper, ``Putting $\{$LTE$\}$ security
  functions to the test: A framework to evaluate implementation correctness,''
  in \emph{10th $\{$USENIX$\}$ Workshop on Offensive Technologies ($\{$WOOT$\}$
  16)}, 2016.

\bibitem{rupprecht2019breaking}
D.~Rupprecht, K.~Kohls, T.~Holz, and C.~P{\"o}pper, ``Breaking lte on layer
  two,'' in \emph{2019 IEEE Symposium on Security and Privacy (SP)}.\hskip 1em
  plus 0.5em minus 0.4em\relax IEEE, 2019, pp. 1121--1136.

\bibitem{rupprecht2020imp4gt}
------, ``Imp4gt: Impersonation attacks in 4g networks,'' in \emph{Symposium on
  Network and Distributed System Security (NDSS). ISOC}, 2020.

\bibitem{sasaki2020security}
T.~Sasaki, S.~Karino, M.~Tani, K.~Nakajima, K.~Tomita, and N.~Yamagaki,
  ``Security architecture for trustworthy systems in 5g era,'' \emph{arXiv
  preprint arXiv:2007.14756}, 2020.

\bibitem{sgandurra2016evolution}
D.~Sgandurra and E.~Lupu, ``Evolution of attacks, threat models, and solutions
  for virtualized systems,'' \emph{ACM Computing Surveys (CSUR)}, vol.~48,
  no.~3, pp. 1--38, 2016.

\bibitem{shakhov2017energy}
V.~Shakhov, I.~Koo, and A.~Rodionov, ``Energy exhaustion attacks in wireless
  networks,'' in \emph{2017 International Multi-Conference on Engineering,
  Computer and Information Sciences (SIBIRCON)}.\hskip 1em plus 0.5em minus
  0.4em\relax IEEE, 2017, pp. 1--3.

\bibitem{shin2013attacking}
S.~Shin and G.~Gu, ``Attacking software-defined networks: A first feasibility
  study,'' in \emph{Proceedings of the second ACM SIGCOMM workshop on Hot
  topics in software defined networking}, 2013, pp. 165--166.

\bibitem{shu2017study}
R.~Shu, X.~Gu, and W.~Enck, ``A study of security vulnerabilities on docker
  hub,'' in \emph{Proceedings of the Seventh ACM on Conference on Data and
  Application Security and Privacy}, 2017, pp. 269--280.

\bibitem{singh2016sql}
N.~Singh, A.~Jangra, U.~Lakhina, and R.~Sharma, ``Sql injection attack
  detection \& prevention over cloud services,'' \emph{International Journal of
  Computer Science and Information Security}, vol.~14, no.~4, p. 256, 2016.

\bibitem{singhal2012security}
A.~Singhal and S.~Singapogu, ``Security ontologies for modeling enterprise
  level risk assessment,'' in \emph{Proceedings of the 2012 Annual Computer
  Security Applications Conference, Orlando, FL, USA}, 2012, pp. 3--7.

\bibitem{sodagari2012efficient}
S.~Sodagari and T.~C. Clancy, ``Efficient jamming attacks on mimo channels,''
  in \emph{2012 IEEE International Conference on Communications (ICC)}.\hskip
  1em plus 0.5em minus 0.4em\relax IEEE, 2012, pp. 852--856.

\bibitem{sodagari2015singularity}
------, ``On singularity attacks in mimo channels,'' \emph{Transactions on
  Emerging Telecommunications Technologies}, vol.~26, no.~3, pp. 482--490,
  2015.

\bibitem{spreitzer2017systematic}
R.~Spreitzer, V.~Moonsamy, T.~Korak, and S.~Mangard, ``Systematic
  classification of side-channel attacks: A case study for mobile devices,''
  \emph{IEEE Communications Surveys \& Tutorials}, vol.~20, no.~1, pp.
  465--488, 2017.

\bibitem{stuttard2011web}
D.~Stuttard and M.~Pinto, \emph{The web application hacker's handbook: Finding
  and exploiting security flaws}.\hskip 1em plus 0.5em minus 0.4em\relax John
  Wiley \& Sons, 2011.

\bibitem{suomalainen2020machine}
J.~Suomalainen, A.~Juhola, S.~Shahabuddin, A.~M{\"a}mmel{\"a}, and I.~Ahmad,
  ``Machine learning threatens 5g security,'' \emph{IEEE Access}, vol.~8, pp.
  190\,822--190\,842, 2020.

\bibitem{tian2017survey}
F.~Tian, P.~Zhang, and Z.~Yan, ``A survey on c-ran security,'' \emph{IEEE
  Access}, vol.~5, pp. 13\,372--13\,386, 2017.

\bibitem{traynor2007attack}
P.~Traynor, P.~McDaniel, T.~La~Porta \emph{et~al.}, ``On attack causality in
  internet-connected cellular networks,'' in \emph{Proceedings of 16th USENIX
  Security Symposium on USENIX Security Symposium}, vol.~21, 2007, pp. 1--21.

\bibitem{tu2015voice}
G.-H. Tu, C.-Y. Li, C.~Peng, and S.~Lu, ``How voice call technology poses
  security threats in 4g lte networks,'' in \emph{2015 IEEE Conference on
  Communications and Network Security (CNS)}.\hskip 1em plus 0.5em minus
  0.4em\relax IEEE, 2015, pp. 442--450.

\bibitem{vadlamani2016jamming}
S.~Vadlamani, B.~Eksioglu, H.~Medal, and A.~Nandi, ``Jamming attacks on
  wireless networks: A taxonomic survey,'' \emph{International Journal of
  Production Economics}, vol. 172, pp. 76--94, 2016.

\bibitem{wang2018stealing}
B.~Wang and N.~Z. Gong, ``Stealing hyperparameters in machine learning,'' in
  \emph{2018 IEEE Symposium on Security and Privacy (SP)}.\hskip 1em plus 0.5em
  minus 0.4em\relax IEEE, 2018, pp. 36--52.

\bibitem{wang2009cooperative}
J.~Wang and A.~L. Swindlehurst, ``Cooperative jamming in mimo ad-hoc
  networks,'' in \emph{2009 Conference record of the forty-third Asilomar
  conference on signals, systems and computers}.\hskip 1em plus 0.5em minus
  0.4em\relax IEEE, 2009, pp. 1719--1723.

\bibitem{yan2015software}
Q.~Yan, F.~R. Yu, Q.~Gong, and J.~Li, ``Software-defined networking (sdn) and
  distributed denial of service (ddos) attacks in cloud computing environments:
  A survey, some research issues, and challenges,'' \emph{IEEE communications
  surveys \& tutorials}, vol.~18, no.~1, pp. 602--622, 2015.

\bibitem{yang2007joint}
L.-L. Yang, ``Joint transmitter-receiver design in tdd multiuser mimo systems:
  An egocentric/altruistic optimization approach,'' in \emph{2007 IEEE 65th
  Vehicular Technology Conference-VTC2007-Spring}.\hskip 1em plus 0.5em minus
  0.4em\relax IEEE, 2007, pp. 2094--2098.

\bibitem{yu2019effects}
C.~Yu and S.~Chen, ``On effects of mobility management signalling based dos
  attacks against lte terminals,'' in \emph{2019 IEEE 38th International
  Performance Computing and Communications Conference (IPCCC)}.\hskip 1em plus
  0.5em minus 0.4em\relax IEEE, 2019, pp. 1--8.

\bibitem{zhang2012deep}
S.~Zhang, ``Deep-diving into an easily-overlooked threat: Inter-vm attacks,''
  \emph{Kansas State University}, 2012.

\bibitem{zhang2018code}
T.~Zhang, G.~Upadhyaya, A.~Reinhardt, H.~Rajan, and M.~Kim, ``Are code examples
  on an online q\&a forum reliable?: a study of api misuse on stack overflow,''
  in \emph{2018 IEEE/ACM 40th International Conference on Software Engineering
  (ICSE)}.\hskip 1em plus 0.5em minus 0.4em\relax IEEE, 2018, pp. 886--896.

\bibitem{zhang2012cross}
Y.~Zhang, A.~Juels, M.~K. Reiter, and T.~Ristenpart, ``Cross-vm side channels
  and their use to extract private keys,'' in \emph{Proceedings of the 2012 ACM
  conference on Computer and communications security}, 2012, pp. 305--316.

\bibitem{zhang2014cross}
------, ``Cross-tenant side-channel attacks in paas clouds,'' in
  \emph{Proceedings of the 2014 ACM SIGSAC Conference on Computer and
  Communications Security}, 2014, pp. 990--1003.

\bibitem{zhang2014iot}
Z.-K. Zhang, M.~C.~Y. Cho, C.-W. Wang, C.-W. Hsu, C.-K. Chen, and S.~Shieh,
  ``Iot security: ongoing challenges and research opportunities,'' in
  \emph{2014 IEEE 7th international conference on service-oriented computing
  and applications}.\hskip 1em plus 0.5em minus 0.4em\relax IEEE, 2014, pp.
  230--234.

\bibitem{zhou2012dissecting}
Y.~Zhou and X.~Jiang, ``Dissecting android malware: Characterization and
  evolution,'' in \emph{2012 IEEE symposium on security and privacy}.\hskip 1em
  plus 0.5em minus 0.4em\relax IEEE, 2012, pp. 95--109.

\bibitem{zou2016survey}
Y.~Zou, J.~Zhu, X.~Wang, and L.~Hanzo, ``A survey on wireless security:
  Technical challenges, recent advances, and future trends,'' \emph{Proceedings
  of the IEEE}, vol. 104, no.~9, pp. 1727--1765, 2016.

\end{thebibliography}
\newpage

\vspace{-60.0pt}
\begin{IEEEbiography}
[{\includegraphics[width=1.1in,height=1.1in]{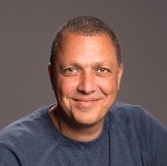}}]{Dudu Mimran}
Tech executive with extensive experience in startups and product building. Core competencies in deep-tech business and product strategy, ideation, and deep tech talent leadership. In recent 20 years innovating in the spaces of cybersecurity, machine learning, big data, and privacy in the enterprise and consumer worlds.
\vspace{-60.0pt}
\end{IEEEbiography}

\begin{IEEEbiography}
[{\includegraphics[width=1.1in,height=1.1in]{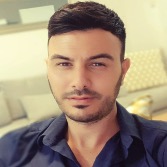}}]{Ron Bitton} is a principal research manager at the Telekom Innovation Laboratories at BGU. His main areas of expertise are the intersection between cybersecurity and machine learning. Ron possesses a B.SC in software engineering, a M.SC in cybersecurity and a Ph.D in Machine Learning all from Ben-Gurion University of the Negev.

\vspace{-60.0pt}
\end{IEEEbiography}

\begin{IEEEbiography}
[{\includegraphics[width=1.1in,height=1.1in]{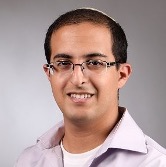}}]{Yehonatan Kfir} is a graduate of the elite Talpiot military academy. He has BSc in Physics and Math from the Hebrew University, MBA from the Technion and MSc in Electrical Engineering from Tel-Aviv University. He holds a PhD in Cyber Security from Bar-Ilan University. Yehonatan research interests is cyber security.

\vspace{-60.0pt}
\end{IEEEbiography}

\begin{IEEEbiography}
[{\includegraphics[width=1.1in,height=1.1in]{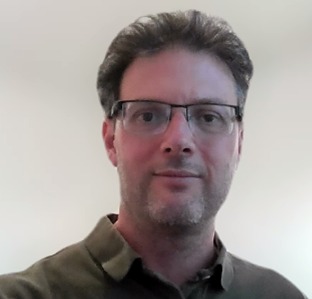}}]{Eitan Klevansky} more than 25 years of experience in leading, design and development of commercial and research projects for startups, large companies and non-profit organizations, in the fields of cyber security, fraud detection, big data and machine learning
\vspace{-60.0pt}
\end{IEEEbiography}

\begin{IEEEbiography}
[{\includegraphics[width=1.1in,height=1.1in]{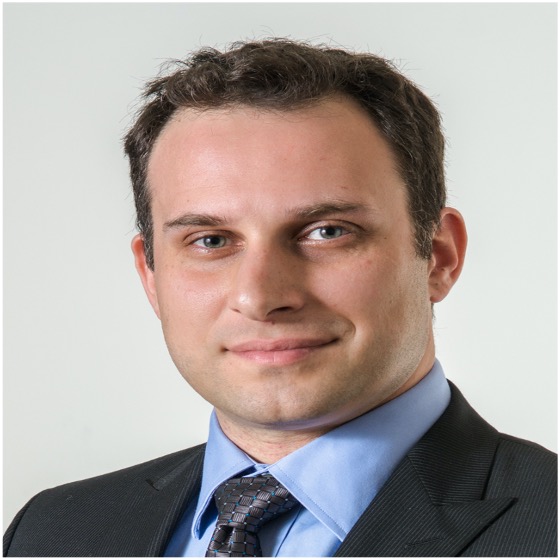}}]{Oleg Brodt} is the R\&D Director of Deutsche Telekom Innovation Labs Israel, focusing on Cyber Security and AI/ML.
He completed engineering studies at the Israeli Air-Force, focusing on networks, computer communications and microelectronics; and gained substantial technological training at an elite IDF unit, where he served as a team leader.
Oleg holds LL.B and LL.M in international Business law and a degree in business and management, from IDC.
\vspace{-60.0pt}
\end{IEEEbiography}

\begin{IEEEbiography}
[{\includegraphics[width=1.1in,height=1.1in]{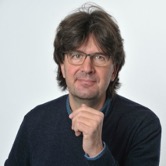}}]{Heiko Lehmann} is Senior Expert for Machine Learning at T-Labs.
He received a PhD in Theoretical Physics from Humboldt University Berlin. Following postdoctoral academic work at Oxford University and the German National Society for Mathematics and Informatics. In 2006, Lehmann joined T-Labs where he took over responsibility for a portfolio of innovation projects focusing on AI/ML and Cybersecurity in modern telecoms networks.
\vspace{-60.0pt}
\end{IEEEbiography}

\begin{IEEEbiography}
[{\includegraphics[width=1.1in,height=1.1in]{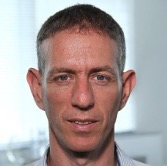}}]{Yuval Elovici} is the director of the Telekom Innovation Labs at BGU, head of the Cyber Security Research Center at BGU, and a professor in the Department of Software and Information Systems Engineering at BGU.
His research interests include computer and network security, and machine learning.
\vspace{-60.0pt}
\end{IEEEbiography}

\begin{IEEEbiography}
[{\includegraphics[width=1.1in,height=1.1in]{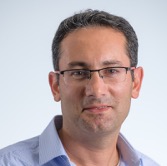}}]{Asaf Shabtai} is a professor in the Department of Software and Information Systems Engineering at Ben-Gurion University of the Negev. 
His main areas of interest are computer and network security, machine learning, and security of IoT, smart mobile devices and operational technology (OT) systems. 
\vspace{-60.0pt}
\end{IEEEbiography}

\end{document}